\documentclass[useAMS,usenatbib]{mn2e}
\usepackage{aas_macros}
\usepackage{amssymb}
\usepackage{amsmath}
\usepackage{times}
\usepackage{graphicx}
\usepackage{multirow}
\usepackage[table]{xcolor}
\usepackage{float}
\usepackage{url}

\newcommand{\owls}{{\small OWLS} }
\newcommand{\urchin}{\textsc{Urchin}}
\newcommand{\sphray}{\textsc{Sphray}}

\newcommand{\nion}[1]{n_{\rm {#1}}}

\newcommand{\xHI}{x}

\newcommand{\NHI}{N_{\rm HI}}
\newcommand{\NH}{N_{\rm H}}

\newcommand{\Inu}{I_{\nu} } 

\newcommand{\nH}{n_{\rm _{H}} }

\newcommand{\nHI}{n_{\rm _{HI}} }

\newcommand{\nHslab}{ \nH = 1.5 \times 10^{-3} {\rm cm^{-3}} }
\newcommand{\nHusphere}{ \nH = 1.5 \times 10^{-3} {\rm cm^{-3}} }

\newcommand{\sigth}{\sigma_{\rm th}}
\newcommand{\nuth}{\nu_{\rm th}}

\newcommand{\signu}{\sigma}

\newcommand{\tauH}{\tau_{\rm H}}

\newcommand{\nel}{n_{\rm e} }

\newcommand{\Msunh}{{\rm M_{\odot}}h^{-1} }

\newcommand{\vect}[1]{\mathbf{#1}}

\newcommand{\ion}[2]{#1{\small\rm{#2}}\relax}

\title[\textsc{Urchin}]{\textsc{Urchin}: 
A Reverse Ray Tracer for Astrophysical Applications}

\author[Altay \& Theuns]
       {Gabriel Altay$^{1}$ and Tom Theuns$^{1,2}$\\
       $^{1}$Institute for Computational Cosmology, Department of
       Physics, University of Durham, South Road, Durham DH1 3LE \\
       $^{2}$Department of Physics, University of Antwerp, Campus
       Groenenborger, Groenenborgerlaan 171, B-2020 Antwerp, Belgium}

\begin{document}

\date{Accepted 201? ???? ??.
      Received 201? ???? ??;
      in original form 2010  xx}

\pagerange{\pageref{firstpage}--\pageref{lastpage}}
\pubyear{201?}
\maketitle

\label{firstpage}

\begin{abstract}
We describe \urchin, a reverse ray tracing radiative transfer scheme optimised to model self-shielding from the post-reionisation ultraviolet (UV) background in cosmological simulations. The reverse ray tracing strategy provides several benefits over forward ray tracing codes including: (1) the preservation of adaptive density field resolution (2) completely uniform sampling of gas elements by rays; (3) the preservation of galilean invariance; (4) the ability to sample the UV background spectrum with hundreds of frequency bins; and (5) exact preservation of the input UV background spectrum and amplitude in optically thin gas. The implementation described here focuses on Smoothed Particle Hydrodynamics (SPH).  However, the method can be applied to any density field representation in which resolution elements admit ray intersection tests and can be associated with optical depths.    We characterise the errors in our implementation in stages beginning with comparison to known analytic solutions and ending with a realistic model of the $z=3$ cosmological UV background incident onto a suite of spherically symmetric models of gaseous galactic halos.  
\end{abstract}

\begin{keywords}
methods: numerical -- 
radiative transfer -- 
intergalactic medium -- 
quasars: absorption lines -- 
diffuse radiation --
ultraviolet: general
\end{keywords}

\section{Introduction}

Cosmological gas dynamics simulations \citep[e.g.][]{Cen_1994,Theuns_1998,Theuns98b,Hernquist_1996}, semi-analytic models \citep[e.g.][]{Bi_1997} and analytic calculations \citep[e.g.][]{Schaye_2001} have enabled us to understand the connection between galaxies, the intergalactic medium (IGM), and the large-scale structure of the Universe.  H{\sc I} Lyman-$\alpha$ absorbers in the spectra of distant quasars are a particularly useful observational probe of this structure.  Models indicate that at redshifts $z=2-4$, the majority of Lyman-$\alpha$ forest absorption lines  (i.e. lines with column densities $\NHI < 10^{17.2}$~cm$^{-2}$) arise in filamentary structures in the IGM with the highest column density lines being associated with the circumgalactic medium of galaxies.  These lines are produced in systems which are highly ionized by the ultraviolet (UV) background \citep[see][for a recent review]{Meiksin_09}.  
An absorber with a column density $\NHI > 10^{17.2}$~cm$^{-2}$ has an optical depth greater than unity for Lyman-Limit photons and is called a Lyman-Limit System (LLS).  Above a column density of $\NHI = 10^{20.3}$~cm$^{-2}$, \lq damping wings\rq\ due to the natural line-broadening of the Lyman-$\alpha$ line are detectable, and the system is called a damped Lyman-$\alpha$ absorber (DLA).  These absorbers probe the interface between the IGM and galaxies as well as the interstellar medium (ISM) of the galaxies themselves. Hydrogen begins to self-shield from the UV background in the LLS column density range, and the reduction in ionising flux plays a major role in setting the ionisation state of these absorbers.  The \urchin\ code described in this paper is designed to model H{\sc I} self-shielding in the LLS and DLA range.

At present, the largest observational catalogues of self-shielded absorbers are produced through semi-automated searches \citep[e.g.][]{Prochaska_04,Noterdaeme_2012} of data from the Sloan Digital Sky Survey (SDSS\footnote{\url{www.sdss3.org}}).  Current and planned expansions to the SDSS such as the Baryon Oscillation Spectroscopic Survey \citep[BOSS,][]{Boss} and BigBOSS \citep{BigBoss} will increase the amount of available data by a factor of ten.  Due to atmospheric absorption of the rest frame Lyman-$\alpha$ transition, ground based surveys for DLAs are limited to redshifts $z>1.6$.  Surveys for LLSs require spectral coverage of the  Lyman-Limit transition for an accurate determination of $\NHI$ and are therefore limited to redshifts of $z>2.5$ when performed from the ground.  The new {\em Cosmic Origins Spectrograph}\footnote{\url{www.stsci.edu/hst/cos}} on the {\em Hubble Space Telescope} provides significant capacity to probe lower redshift systems \citep[e.g.][]{Battisti_12} while the {\em Advanced Camera for Surveys} and {\em Wide Field Camera 3} have recently been used to complete a survey for LLSs in the redshift range $1.0 < z < 2.6 $ \citep{Omeara_12}.

In addition to Lyman series transitions, post-reionisation neutral hydrogen can also be effectively probed using the 21-cm emission line.  The most recent determination of the local \ion{H}{I} mass function is from the Arecibo Legacy Fast ALFA (ALFALFA) survey \citep{Martin_10} which will have detected $\approx 3 \times 10^4$ galaxies in \ion{H}{I} 21-cm out to $z = 0.06$ when it is complete.  The Square Kilometer Array (SKA\footnote{\url{www.skatelescope.org}}) represents the long term future for this type of radio astronomy however construction will not begin for several years.  In preparation of this, a host of pathfinding telescopes ( ASKAP\footnote{\url{www.atnf.csiro.au/projects/askap} }, MeerKAT\footnote{\url{www.ska.ac.za/meerkat}}, WSRT\footnote{\url{www.astron.nl/radio-observatory/astronomers/wsrt-astronomers}}, EVLA\footnote{\url{https://science.nrao.edu/facilities/evla}} ) will soon make 21-cm emission surveys as data rich as their optical and near-infrared counterparts.  In addition, pilot surveys for 21-cm absorption in the spectra of radio bright sources have shown potential \citep[e.g.][]{Gupta_10, Darling_2011} and at least two of these pathfinders (ASKAP and MeerKAT) will also perform large, blind, absorption surveys. 

The combined output from these surveys will provide transformative information on the gas content of galaxies and their modes of accretion.  It will also generate samples that trace the large scale structure of the Universe with different biases than those of optically selected samples.  A prerequisite for making model predictions of hydrogen emission or absorption is the accurate calculation of the distribution of neutral hydrogen 
\citep[e.g.][]{Duffy_12,VandeVoort_12}.  Methods that accomplish this task with a minimum of free parameters will be able to take full advantage of observational data.  These issues motivate many of the design choices for \urchin\ .

The standard approach of treating the post-reionization UV background in cosmological simulations is to impose a spatially uniform but time varying radiation field, and calculate the \ion{H}{1} fraction in the optically thin limit. Pioneering work on modeling self-shielding in gas dynamics simulations was done by \cite{Katz_1996} and \cite{Haehnelt_1998} by post-processing column density maps.  More recent theoretical work has incorporated radiative transfer through 3-D density fields to calculate the attenuation of the UV background in dense gas \citep{Razoumov_2006, Kohler_07, Pontzen_2008, Altay_2011, McQuinn_2011, Yajima_11, Fumagalli_11, Cen_12, Erkal_2012, Rahmati_13_evo, Rahmati_13_str}.

In this paper we present and test \urchin, a reverse ray tracing scheme designed to calculate self-shielding corrections in the post-reionisation Universe.  The code can be applied to any density field representation (e.g. particles, adaptive grids, unstructured meshes) in which the resolution elements can be associated with optical depths and subjected to ray intersection tests.  The main benefits of \urchin\ are: (1) preservation of the adaptive density field resolution present in many gas dynamics codes; (2) uniform sampling of gas resolution elements with rays; (3) preservation of galilean invariance; (4) high spectral resolution; and (5) preservation of the standard uniform UV background in optically thin gas.  The format of this paper is as follows.  In \S2 we introduce our notation and review some basic physics related to radiative transfer.  In \S3  we give a general description of our reverse ray tracing approach and place it in context by comparing it to alternative approaches.  In \S4 we discuss the details of our implementation using smoothed particle hydrodynamics (SPH) density fields. In \S5 we present the results of tests meant to validate \urchin\ and in \S6 we discuss the results, suggest improvements for future versions of the code, and conclude.

\section{Definitions and Basic Physics}
\label{definitions}

In this section, we define our notation and review some of the relevant physics.  All simulations discussed in this work utilize a cubic simulation volume.  For brevity, we will refer to any of these simulation volumes as boxes and their six faces as walls.   When refering to distances, we will distinguish between proper and comoving measures using the prefixes \lq p \rq and \lq c \rq (e.g. pkpc, cMpc).  In this work, we consider only hydrogen and leave the inclusion of other elements, particularly helium, to future work. For our description of the radiation field, we adopt the notation of \cite{Rybicki_86}. 

The specific intensity, $I_{\nu} \equiv dE / dA \, d\Omega \, dt \, d\nu$, fully characterises a radiation field and is defined as the energy $dE$ passing through an area element $dA$ into a solid angle element $d\Omega$ in time $dt$ due to photons with frequency between $\nu$ and $\nu+d\nu$.  Several useful characterisations of the radiation field can be expressed as integrals over this quantity.  Considering photons with frequency $\nuth < \nu < q \nuth$, and an optically thin medium, we can write the number density of hydrogen ionising photons and the photoionisation rate at the point $\vect{x}$ respectively as:
\begin{eqnarray}
  \label{eq:ngamma} 
  n_{\gamma} (\vect{x}) &=&
  \frac{1}{c} 
  \oint 
  \int_{\nuth}^{q \nuth}  
  \frac{\Inu}{h \nu} 
  d\nu \, 
  d\Omega,\\
  \label{eq:gamma}
  \Gamma (\vect{x}) &=&
  \oint 
  \int_{\nuth}^{q \nuth} 
  \frac{\Inu \signu}{h \nu} 
  d\nu \,
  d\Omega\,
\end{eqnarray}
where $h \nuth = E_{\rm th}$ and $\sigma$ are the ionisation energy and photoionisation cross-section of hydrogen. In the case of a medium with finite opacity, the above quantities can be written in terms of the optically thin value of $\Inu$ by making the replacement $\Inu \rightarrow \Inu \exp \left[ -\tau(\nu,\Omega) \right]$ where $\tau$ is the optical depth between the sources producing $\Inu$ and $\vect{x}$.  

The frequency averaged (\lq grey\lq) photoionisation cross section is defined as $\sigma_{\rm grey}\equiv \Gamma\,(c n_{\gamma})^{-1}$.  Under the grey approximation, every polychromatic spectrum (characterised by $n_{\gamma}$ and $\Gamma$) corresponds to an equivalent monochromatic spectrum with the same $n_{\gamma}$ and $\Gamma$.    The flux of the monochromatic spectrum is fixed by $n_{\gamma}$ and the energy of the photons in the monochromatic spectrum is  $E_{\rm grey} = h \nu_{\rm grey}$ with $\nu_{\rm grey}$ implicitly defined by $\sigma(\nu_{\rm grey}) = \sigma_{\rm grey} $.  

\section{Reverse ray tracing Method}
\label{sect:method}

\urchin\ is designed to efficiently model the residual neutral hydrogen in the post-reionisation universe.  In this section, we describe the algorithms used in \urchin\ and conceptual departures from alternative radiative transfer codes.  We begin with a brief description of the standard treatment of the UV background in cosmological gas dynamics simulations.

\subsection{Standard Treatment of Post-Reionisation UV Background} 

In the post-reionisation Universe, cosmic hydrogen is kept highly ionised by a pervasive UV background \citep{Gunn_65} produced by galaxies and quasars. Quantitatively, the volume averaged neutral fraction $\xHI = \nHI/\nH$ for redshifts $z \le 6$ is on the order of $10^{-4}$ \citep{Fan06,Becker07}.  A rapid transition to higher neutral fractions signals the end of reionisation, evidence for which has recently been observed in the form of a damping wing in the spectrum of a $z \sim 7$ quasar \citep{Mortlock_11}.  At lower redshifts, the UV background determines both the ionisation state and temperature of gas in the IGM and sets the rate at which denser gas can cool, accrete onto small galaxies, and form stars \citep{Efstathiou_1992,Okamoto_08}.  

The most widely used treatment of the post-reionisation UV background in cosmological simulations, is based on three approximations: (1) optically thin gas; (2) a spatially uniform UV background; and (3) photo/collisional ionisation equilibrium.  These approximations are valid for the majority of cosmic gas, i.e. for highly ionised hydrogen; however, they break down for the majority of gas observable in \ion{H}{I} surveys.  The approximation of optically thin gas begins to break down in absorption systems that probe accretion (LLSs) from the IGM onto galaxies and completely fails in regions of significant self-shielding (DLAs) where the strongest \ion{H}{I} signals arise.  This approximation is the most important to remedy for \ion{H}{I} surveys.  The second approximation involves disregarding large scale gradients in the UV background as well as point sources.  These fluctuations can have an effect on absorber statistics \citep[e.g.][]{Rahmati_13, Yajima_11, Croft_04} but the magnitude is not as large as that due to self-shielding.  A notable exception is absorbers in the proximity zones of bright sources \citep{Schaye_06}.  The third approximation involving photo/collisional equilibrium likely holds for dense gas where \ion{H}{I} is self-shielded and recombination times are short compared to UV background variability, but will break down near variable sources and in less dense gas.  Relaxing these approximations is an efficient way to improve model predictions for \ion{H}{I} surveys.  In this release of \urchin\ we focus on self-shielding.

\subsection{Reverse Ray Tracing - Motivation}

Numerical techniques for continuum radiative transfer have been developed based around ray tracing methods, \citep[e.g.][]{Nakamoto_01, Maselli_03, Razoumov_05, Susa_06, Whalen_06, Altay_08}, the closely related method of characteristics \citep[e.g.][]{Mellema_06,Rijkhorst_06}, angular moments of the radiative transfer equation \citep[e.g.][]{Gnedin_01,Aubert_08,Petkova_09,Finlator_09}, and transport on unstructured meshes \citep[e.g.][]{Pawlik_08,Pawlik_11,Paardekooper_10}.  A detailed comparison between many codes currently in use is documented in the Cosmological Radiative Transfer Comparison Project \citep{Iliev_06,Iliev_09}.  

The above methods are all based on following radiation from its source to the point where it is absorbed or scattered.  When dealing with the post-reionisation UV background, the goal is to build up a radiation field that is known from both theoretical and observation studies to be mostly uniform.  In these methods, the UV background field at a given point is the sum of radiation that has been transported from all sources being considered.  However, in many methods this is true only in a statistical sense.  For example, in Monte Carlo ray tracers such as \sphray, resolution elements are updated whenever a ray intersects them.  The ray only carries information about the flux from one source, but if enough rays are traced, each resolution element will be updated by the rays from many sources.   In \urchin\ we attempt a different type of solution to this problem.  As opposed to building up a mostly uniform UV background from multiple sources, we begin with the standard approximations described above (optically thin gas, uniform field, photo-collisional equilibrium) and then calculate deviations from it.  For the majority of cosmic gas, the optically thin photoionisation rate from this uniform field, $\Gamma^{\rm thin}$, is in fact a good approximation.  The current version of \urchin\ relaxes the optically thin approximation from the standard treatment by attenuating this uniform radiation field in denser regions that begin to self-shield.  The fraction of gas (by mass or volume) where this correction is necessary is guaranteed to be small by the nature of the problem allowing us to concentrate the available computational resources where they are needed.  In future versions of the code we will relax the second and third approximations in the standard treatment by adding large scale gradients, proximity regions, and considering non-equilibrium effects.  Currently \urchin\ operates on static density fields as a post-processing step, but in principle could be coupled to existing gas dynamics codes in a straight-forward way \cite[e.g. in a framework such as described in][]{Zwart_09}.

\subsection{Reverse Ray Tracing - Algorithm}

Consider a density field discretised into resolution elements labeled $i=1,\cdots,N$.  We will refer to these resolution elements as particles; however, our reverse ray tracing technique can be applied to any density field representation in which the resolution elements can be associated with optical depths and subjected to ray intersection tests.  From each particle, we cast $N_{\rm ray}$ ray segments out to a distance $l_{\rm ray}$, in directions that uniformly cover the solid angle around the particle.  We use the HEALPix algorithm \citep{2005ApJ...622..759G} to determine ray directions.  We then calculate the optical depth $\tau_k$ along each of these ray segments and sum over all rays to obtain an estimate of the self-shielded photoionisation rate, $\Gamma^{\rm shld} \le \Gamma^{\rm thin}$, at the location of each particle,
\begin{eqnarray}
  \frac{\Gamma^{\rm shld}}{\Gamma^{\rm thin}} &\equiv& \exp(-\tau_{\rm eff}) \nonumber \\
  &=& \frac{1}{\Gamma^{\rm thin}} \frac{4\pi}{N_{\rm ray}} \sum_{k=1}^{\rm N_{\rm ray}} 
  \int_{\nuth}^{q \nuth} \frac{\Inu \sigma}{h\nu} \exp(-\tau_k) d\nu \,.
  \label{eq:gamma_shld}
\end{eqnarray}
For most particles the optically thin approximation is very good and the effective optical depth will be small, $\tau_{\rm eff} \ll 1$, along all $N_{\rm ray}$ directions.  However, for the small fraction of particles that dominate the \ion{H}{I} abundance, the effective optical depth will be large ($\tau_{\rm eff} \gg 1$).  We then use $\Gamma^{\rm shld}$ in an analytic solution for the equilibrium neutral fraction $x$ (Eq.~\ref{eq:xroots} in the Appendix) to update the ionisation state of the particle.  An altered neutral fraction for one particle leads to altered optical depths along ray segments that pass through it, and hence we iterate this procedure until the neutral fraction converges for all particles.  This typically happens in tens of iterations with those few particles at the threshold between optically thin and optically thick converging last.  $l_{\rm ray}$ and $N_{\rm ray}$ are numerical parameters of the scheme, and we have found that in fully cosmological runs at $z\sim 3$, values of $l_{\rm ray} = 100$~pkpc and $N_{\rm ray} = 12$ lead to converged column density distribution functions.

\subsection{Reverse Ray Tracing - Advantages}

Reverse ray tracing presents some important advantages in the post-reionization regime, related to the dramatic differences in the character of the radiation field during and after reionization.   In the following sub-sections, we focus on questions that typically occur in ray tracing schemes attempting to model the post-reionization cosmological UV background: 1) Where should rays originate and terminate?  2) How should rays be traced to uniformly sample the gas elements? 3) How much spectral resolution can be attained? and 4) What optimizations can be applied?  We will show that answers to these questions are simpler in reverse ray tracing schemes than in forward versions.



\subsubsection{Where Should Rays Originate?}

In forward ray tracing schemes, the UV background at a given point in space is the result of transporting photons along rays originating at sources of radiation.  These sources can be divided into two categories: those inside the box (internal sources) and those outside the box (external sources).  Modelling the UV background from first principles using only internal sources imposes stringent conditions on the size of the box.  A reasonable absolute minimum scale is one mean free path at the Lyman limit, $\lambda^{\rm mfp}_{912}$.  \cite{Prochaska_09} find that this mean free path depends on redshift $z$ approximately as $\lambda(z)\approx \left[(48\pm2.1)-(38.0\pm5.3)(z-3.6)\right]~h_{72}^{-1}$~pMpc for $3.6 < z < 4.3$.  At $z=3.6$  this is larger than the majority of cosmological gas dynamics simulations able to resolve any galaxy formation processes and $\lambda^{\rm mfp}_{912}$ becomes larger at lower redshifts.  For this reason, forward ray tracing schemes inevitably must rely on tracing many rays from external sources in addition to internal ones. 

The contribution from external sources is usually modeled by casting rays inward from the walls.  These rays are then followed until a given fraction of their initial photon content is absorbed.  The most straight forward method of choosing ray origins is a pseudo-random sampling of points on the walls.  The appropriate flux is usually determined by requiring that the photoionisation rate in optically thin gas match an input value.  This method has two drawbacks.  The first, discussed in the next sub-section, is the difficulty in uniformly sampling an adaptive density field.  The second is the artificial gradient in photoionisation rate that will develop between regions near the centre of the box, where the background rays will be most attenuated, and regions near the walls, where the background will be at its optically thin strength.  This is what we term the galilean invariance problem.  This adds ambiguity to any calibration of flux from the walls and makes the calibration dependent on the box size.  In general, tracing rays from the walls leads unavoidably to a loss of symmetry in the UV background and difficulty in flux calibration.  
These problems can be alleviated somewhat by tracing rays from randomly selected under-dense regions within the box \citep{Maselli_05}, but not eliminated entirely.  Some authors \citep[e.g.][]{McQuinn_2011} have circumvented this problem by excising small regions around halos and ray tracing them individually however this neglects filamentary gas between halos.    

In a reverse ray tracing method, these problems are not present.  All symmetries of the UV background are maintained and the photoionisation rate at a given particle position is independent of any properties of the box.  In addition, the intensity of the UV background in optically thin gas is equal to its value in the standard approximation (optically thin gas, uniform field, photo-collisional equilibrium) by construction.  This conveniently allows for direct comparisons between results with and without radiative transfer with no need for calibrating fluxes. 

\subsubsection{The Uniform Sampling Problem}

Modern gas dynamical simulations discretise density fields with adaptive resolution elements.  This typically leads to better spatial resolution (i.e. smaller resolution elements) in regions of higher gas density.  A radiative transfer approach which relies on rays uniformly cast from the simulation walls  leads to poor sampling of the resolution elements.  Specifically, the overdense self-shielded regions where radiative transfer is most important are undersampled, and the underdense optically thin regions where radiative transfer is unnecessary are oversampled.  A scheme in which rays split and merge \citep{Wise_11} can maintain a constant number of rays intersecting each resolution element but comes at the cost of increased overhead.  On the other hand, the reverse ray-tracing algorithm samples each particle with the same number of rays by construction.  This helps focus computational power where it is needed. 

\subsubsection{Spectral Resolution} 

In a forward ray tracing scheme, the radiation field at a particle position is the summation of contributions from different rays.  For monochromatic spectra, all attenuation information can be characterised by the number of photons in a ray.  To accurately handle multi-frequency spectra, one must substantially increase the number of monochromatic rays traced or carry spectral information along with each ray.  Similarly for  moment methods, each frequency group must be treated separately by the solver. 
{ 
The commonly used grey approximation, i.e. an optimal monochromatic choice, can lead to order of magnitude errors in the neutral fraction (see Figure~\ref{fig:slab} below).  Additionally, \cite{Mirocha_12} have shown that at least four frequency bins between 13.6 and 100 eV are required to obtain converged results when modelling neutral fractions in \ion{H}{II} regions produced by smooth spectra such as thermal emission and power laws.  To include the effects of sharp spectral features, for example the Helium Lyman-$\alpha$ recombination line present in the UV background spectrum of \cite{HM_12} (see Figure~\ref{test_one_plt}), more frequency bins are required.  
}

In reverse ray tracing, the rays simply sample the optical depth along a particular direction.  Subsequently, arbitrarily complex input spectra can be numerically integrated over frequency to determine a shielded photoionisation rate, and update the neutral fraction of a particle based on full knowledge of the amplitude and spectral shape of the local radiation field.  When using 100 frequency bins in \urchin, the numerical integration over the UV background spectrum consumes a negligible fraction of the computing time.  { This level of spectral resolution is comparable to the resolution with which modern spectral models are defined.  For reference, the recent model spectrum of \cite{HM_12} contains 150 samples between one and ten Rydbergs.   }  This allows for studies of spectral hardness and eases the inclusion of sharp features such as recombination lines into the UV background spectrum.  As an added bonus, knowledge of the optical depth in all directions around a particle during the update allows one to choose the appropriate \lq case A\rq\ (recombinations to all levels) or \lq case B\rq\ (recombinations to all but the ground state) recombination rates in the on-the-spot approximation.

\subsubsection{Optimisations}

The reverse ray tracing approach also lends itself to some important optimisations that are not easy to implement in forward ray tracing techniques.  The most expedient is simply avoiding radiative transfer where it is unnecessary.  As discussed above, particles that are not self-shielded and not in proximity zones are treated correctly in the uniform and optically thin limit. When looping over the particles in reverse ray tracing schemes, those that satisfy this criterion can be skipped.  Identifying these particles is implementation dependent; however, in cosmological simulations, the majority of post-reionization gas does not need to participate in radiative transfer.  We will discuss our technique for identifying these particles in the next section. 

There is also a sense of ray locality and ray independence inherent in the method that can be useful in optimisations.  Any sub-volume in a box can be treated independent of any other as long as a buffer of $l_{\rm ray}$ is included around the sub-volumes.  In addition, each ray can be made independent during a single iteration by waiting to update the neutral fractions of particles until shielded photoionisation rates have been calculated for all of them.  This strategy may increase the number of iterations necessary for convergence but would make each iteration much faster.


\section{Implementation Details}

\urchin\ applies a specified radiation field to a given density field to determine the properties of self-shielded regions. Spatially uniform models of the cosmological UV background which provide $\Inu(z)$, are publically available \citep[for example][]{HM_12}\footnote{\url{http://www.ucolick.org/~pmadau/CUBA}}.  The main parameters in \urchin\ are the frequency range of ionising photons included in the calculation, the number of rays casts per particle, $N_{\rm ray}$, their proper length, $l_{\rm ray}$, the criterion for choosing between type A and type B recombination rates, and the condition for convergence $\delta x_{\rm tol}$.  The choice of spectrum and frequency range determines the optically thin photoionisation rate, $\Gamma^{\rm thin}$, from 
Eq.~(\ref{eq:gamma}). Our default runs use $\nuth < \nu < 4 \nuth$, $N_{\rm ray}=12$ and $l_{\rm ray}=100$~pkpc, which give converged numerical results at redshift $z=3$. In addition, we use case B recombination rates for particles with $\tau_{\rm eff} \ge 1$.

Initially the neutral fraction $\xHI$ of each particle is set to its optically thin value,  
$\xHI_{\rm thin} = \nion{HI}^{\rm thin} / \nion{H}$.  
This allows for the calculation of the \ion{H}{I} column density along each ray cast from a particle. 
Next we calculate a shielded photoionisation rate, $\Gamma^{\rm shld}$, for each particle and an effective optical depth, $\tau_{\rm eff}$ using Eq.~(\ref{eq:gamma_shld}). Particles with $\tau_{\rm eff} < \tau_{\rm eff}^{\rm skip}$ maintain $\xHI = \xHI_{\rm thin}$ for all subsequent iterations while all other particles are updated each iteration.  We continue to loop over particles which initially had $\tau_{\rm eff} > \tau_{\rm eff}^{\rm skip}$ until convergence in the neutral fraction, $|\delta x|/x < \delta x_{\rm tol}$.
Typical values for these parameters are $\tau_{\rm eff}^{\rm skip}=10^{-4}$, $\delta x_{\rm tol}=10^{-3}$, $l_{\rm ray}=100$~pkpc and $N_{\rm ray}=12$.
For these choices we find that more than 99 percent of particles have converged neutral fractions after five iterations in a cosmological run, with the remaining 1 percent requiring tens of iterations.  In our current implementation, we update the neutral fraction of each particle as soon as its self-shielded photoionisation rate has been computed. This update scheme works well provided the list of particles is traversed from high to low density, however, there is no fundamental restriction on when the particles should be updated.  Implementations which calculate $\Gamma^{\rm shld}$ for each particle before performing any updates of the neutral fractions are independent of the order in which the particles are looped over, which could be useful in parallel strategies.

\subsection{Reverse Ray Tracing in SPH}

Although \urchin\ can be applied to many types of density field discretisations (particles, grids, unstructured meshes), in this section  we discuss our implementation in Smoothed Particle Hydrodynamics \cite[SPH,][]{Gingold_77, Lucy_77}.

\subsubsection{Column Densities in SPH}
Our calculation of SPH column densities makes use of several improvements over the algorithm described in \cite{Altay_08}. In the SPH formalism the number density of \ion{H}{I} at any location $\vect{r}$ can be calculated using the scatter approach as
\begin{equation}
\nion{HI}(\vect{r}) = \sum_i \frac{m_i x_i}{m_p} \, W (q_i)\,,
\label{eq:nhi}
\end{equation}
where $m_p$ is the mass of a proton, $m_i$ is the mass of particle $i$ in hydrogen, $x_i$ is its neutral fraction, $q_i\equiv |\vect{r}-\vect{r}_i|/h_i$ is the distance between the particle and and the point $\vect{r}$ in units of the particle's smoothing length $h_i$, and $W$ is the SPH smoothing kernel.  The column density through an SPH distribution along a path $\vect{r}(l)$ parameterised by $0<l<L$ can then be written as:

\begin{eqnarray}
\NHI &=& \int_0^L \nion{HI}(\vect{r}) \, dl
= \int_0^L \sum_i \frac{m_i x_i}{m_p} \, W (r_{il}, h_i) \, dl \nonumber \\
 &=& 
  \sum_i \frac{m_i x_i}{m_p} \int_0^L W (q_{il}) \, dl\,,
\label{eq:colden}
\end{eqnarray}
where the summation is over particles with smoothing volumes intersected by the path $\vect{r}(l)$ and the subscripts in the variable $q_{il}$ indicate that it is a function of the summation index $i$ and distance along the path $l$.  In this way, the calculation of optical depths is reduced to calculating which particles are intersected by a ray and line integrals through the smoothing kernel $W$.

The Gaussian function has many properties that make it a natural choice for the smoothing kernel, however, its lack of compact support leads to an impractical sum over all particles in the simulation volume.  To remedy this, many SPH codes make use of spline functions with an approximately Gaussian shape, such as the $M_4$ cubic spline \citep{MonLat85},

\begin{equation}
M_4(q) = \frac{8}{\pi h^3} \left\{
     \begin{array}{lr}
       1 - 6 q^2 + 6 q^3 & 
       {\rm for} \quad  0 \le q \le \frac{1}{2} \\
       2 (1 - q)^3 & 
       {\rm for} \quad  \frac{1}{2} < q \le 1 \\
       0 & {\rm otherwise\,.}
     \end{array}
   \right.
   \label{eq:spline}
\end{equation}
In \urchin\ we use a truncated Gaussian kernel that allows us to obtain line integrals at a given impact parameter analytically.  A normalised Gaussian function centred at the origin is given by
\begin{equation}
  G(r,\sigma^2) = \frac{\exp(-A^2 r^2)}{(2 \pi \sigma^2)^{3/2}}\,,
\end{equation}
where $A^2 \equiv (2 \sigma^2)^{-1}$ and $\sigma^2$ is the variance. We truncate
$G$ at $r=h$ and determine $\sigma(h)$ using the equation $G(0,\sigma^2) = M_4(0,h)$ but demand the kernel satisfies the condition $\int_0^h 4 \pi r^2 G_t dr = 1$ to obtain the normalization $\mathcal{N}$.
\begin{equation}
G_t(r,\sigma) = \mathcal{N}
\left\{
\begin{array}{lr}
 \exp(-A^2r^2) &\quad {\rm for}\quad {r\le h}\\
 0 & \quad{\rm otherwise}
 \end{array}
 \right.
 \label{eq:Gt}
\end{equation}
where 
\begin{eqnarray}
  \sigma^2 &=& \frac{h^2}{8 \pi^{1/3}}\nonumber\\
  A^2 &=& \frac{4 \pi^{1/3}}{h^2}\nonumber\\
  \mathcal{N} &=& \frac{8}{\pi h^3}
  \left[ 
    {\rm erf(t)} - \frac{2 t \exp(-t^2)}{\sqrt{\pi}}  
    \right]^{-1}\nonumber\\
  t&=& 2 \pi^{1/6}\,,
\end{eqnarray}
and the error function is defined as,
\begin{equation}
  {\rm erf}(t) = \frac{2}{\sqrt{\pi}}
  \int_0^t \exp(-t^2) dt\,.
\end{equation}
The column density through such a Gaussian kernel at impact parameter $b$, between the limits $z_1$ and $z_2$ is then
\begin{equation}
  I(h,b,z_1,z_2) = \frac{\mathcal{N} \sqrt{\pi} \exp(-A^2 b^2)}{2A} 
  \left[ {\rm erf}(A z_2) - {\rm erf}(A z_1) \right]\,.
\end{equation}
We use this kernel to calculate the \ion{H}{I} column densities along each ray using Eq.~(\ref{eq:colden}).  Incidentally, the fact that the Gaussian kernel can be decomposed into three 1-D functions makes it useful for smoothing SPH particles onto 2-D or 3-D grids. \\

\subsubsection{Self-contribution to self-shielding in SPH}
\label{sect:partupdate}
When passing from a continuous density field representation to a discrete one, Eq. (\ref{eq:xroots}) for the equilibrium neutral fraction at the point $\vect{r}$, $x = x(\vect{r},\Gamma,\nH,T,y)$ becomes partially implicit: in the discrete case, $\Gamma$ has a dependence on $x$ through the particle's own contribution to the optical depth along a ray. 
This is a generic problem of discretisation and has been discussed previously in \cite{Abel_99} and \cite{Mellema_06}.  

\begin{figure}
\begin{center}
\includegraphics[width=0.90\columnwidth]{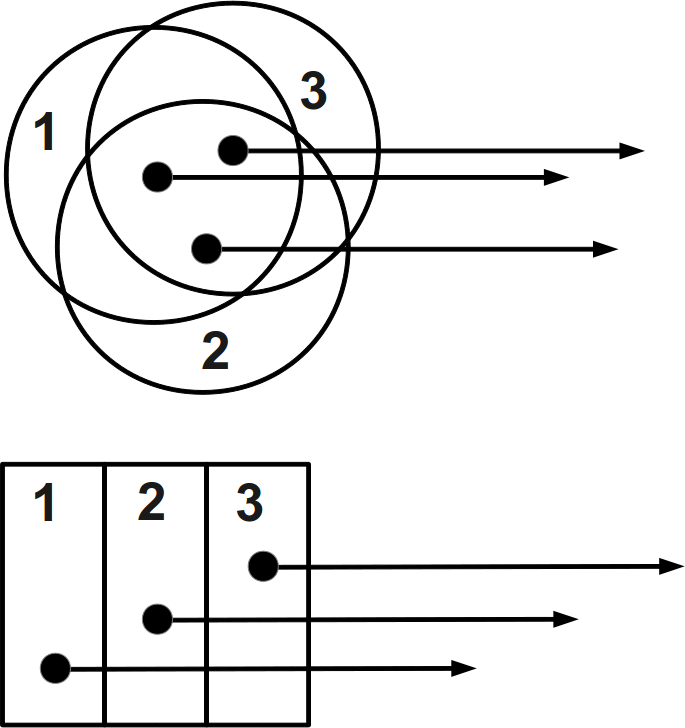}
\end{center}
\caption{An illustration of two density field discretisation methods.  The top panel shows a multi-valued relationship between points in space and resolution elements (SPH).  The bottom panel shows a single valued relationship between points in space and resolution elements (rectilinear).  We show a single ray of length $l_{\rm ray}$ traced from each resolution element.  In the SPH case, a change in the neutral fraction of any element changes the optical depth calculated along all three rays.  In the rectilinear case, changing the neutral fraction of element 1 (for example) leaves the optical depth calculated along the rays from elements 2 and 3 unchanged.}
\label{fig:sphmulti}
\end{figure}

However when the density field is represented with SPH particles, further complications arise due to the weighted sum over neighbours.  For example the \ion{H}{I} density at location $\vect{r}$ depends on the neutral fractions of all particles that appear in the sum of Eq.~(\ref{eq:nhi}). This makes a straightforward scheme that updates neutral fractions based on  $\Gamma^{\rm shld}$ calculated at the center of an SPH particle unstable.
The reason for this instability is illustrated in the top part of Figure~\ref{fig:sphmulti}.  Consider three SPH particles which are initially optically thin but are located in a region that will eventually become self-shield. Rays traced outward (in any direction) from each particle will probe optical depth contributed by all three.  This will in turn cause the neutral fraction of each particle to be increased.  In the next iteration, each ray will find increased optical depth and each neutral fraction will increase again.  This process will continue and eventually spread to adjacent particles causing unphysical growth of neutral regions. The cause of this instability is the multi-valued relationship between points in space and resolution elements in SPH.  
As a counterexample, consider the uniform grid density field represented in the bottom part of Figure \ref{fig:sphmulti}.  This discretisation has a single-valued relationship between points in space and resolution elements.  In this case, changes in the neutral fraction of each element do not necessarily change the optical depth encountered by each ray.  For example, an increase in the neutral fraction of element 1 has no effect on the optical depth encountered by rays traced from elements 2 or 3.  This prevents the instability from occuring. 

We resolved this numerical artefact in SPH as follows. Particles intersected by each ray are split into two distinct groups labeled {\em near} and {\em far}, with $\tau = \tau^{\rm near} + \tau^{\rm far}$.  Near particles of particle $i$ are those that contribute to $i$'s (neutral) density in Eq.~(\ref{eq:nhi}), {\em i.e.} the particle's neighbours. All other particles that contribute to the sum in Eq.~(\ref{eq:colden}) are labelled \lq far\rq. The numerical instability only involves near particles, therefore we can treat the {\em far} particles with the algorithm described in \S3. We model
the near particles 
as a uniform slab with the same temperature $T$ and density $\nion{H}$ as particle $i$.  The thickness of the slab is quantified by calculating the total hydrogen optical depth of the {\em near} particles at the Lyman Limit defined as  $\tau^{\rm near}_{\rm H} \equiv \NH^{\rm near} \sigma_{\rm th}$.  The column density $\NH^{\rm near}$ is calculated as in Eq.~(\ref{eq:colden}) except all of the neutral fractions are set to unity.  

The radiation incident onto the slab is determined by the user supplied spectrum and the optical depth through the {\em far} particles, $\Inu \exp(-\tau^{\rm far})$.  For each ray $k=1,\cdots,N_{\rm ray}$ this incoming flux provides a photoionisation rate at the surface of the slab of $\Gamma^{\rm far}_k$.  We then solve for the ionisation structure in the slab and associate the photoionisation rate at the bottom, $\Gamma^{\rm shld}_k$, with the contribution from ray $k$ to the total shielded photoionisation rate $\Gamma^{\rm shld}$ of particle $i$.  In the case of monochromatic radiation, there is an analytic solution for photoionisation rate as a function of depth into the slab which can be used to determine $\Gamma^{\rm shld}_k$ (see Appendix \ref{sect:slabsoltn}).  For polychromatic spectra we tabulate the solution as a function of four variables, 
$\{ 
\Gamma^{\rm far} / \nH ,\, 
\tau^{\rm far} ,\, 
T ,\, 
\tau^{\rm near}_{\rm H} \}$
\footnote{The optical depth $\tau^{\rm far} = \NHI^{\rm far} \sigma$ is a function of frequency and not a scalar like the other parameters, however, the full shape of $\tau^{\rm far}(\nu)$ is determined by a single evaluation, for example $\tau^{\rm far}(\nuth)$. }.  

\begin{figure}
\begin{center}
\includegraphics[width=0.90\columnwidth]{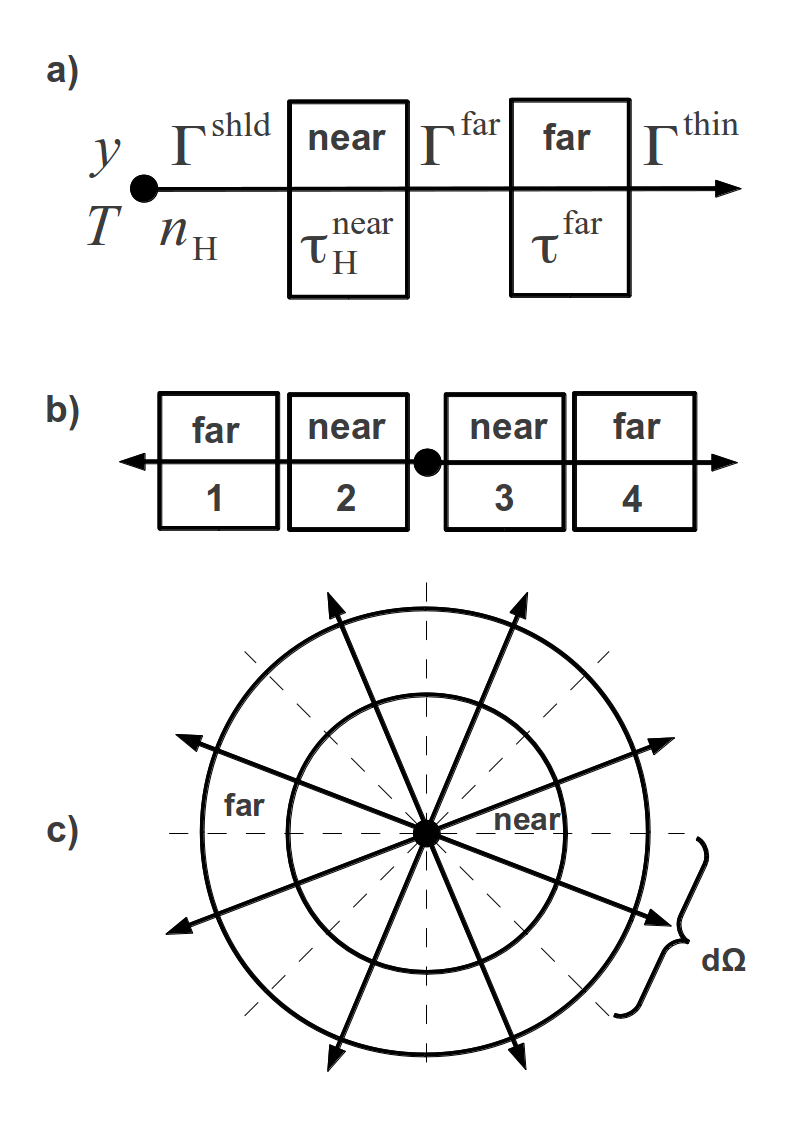}
\end{center}
\caption{An illustration of our reverse ray tracing technique in SPH density fields.  Panel a) shows a typical situation in Test 2 with a single ray being traced from a particle with density $\nH$ and temperature $T$.  The {\em far} slab on the right indicates the optical depth $\tau^{\rm far}$ due to {\em far} particles which attenuate the optically thin radiation field and determine $\Gamma^{\rm far}$ (Eq.~\ref{eq:gammafar}).  This, in turn, determines the radiation incident on the slab used to model the {\em near} particles.  The density and temperature of the {\em near} slab are determined by the particle being updated ($\nH$,$T$) but its thickness $\tauH^{\rm near} = \NH^{\rm near} \sigma_{\rm th}$ is fixed by all {\em near} particles.  The solution $\mathcal{G}$ is then used to determine the photoionisation rate $\Gamma^{\rm shld}$ at the bottom of the slab independent of the neutral fractions of any of the {\em near} particles (Eq.~\ref{eq:gammashldk}).  The variables $T$,$\nH$,$y$, and $\Gamma^{\rm shld}$ are then used in an analytic solution (Eq.~\ref{eq:xroots}) to determine the updated $x$ for the particle.  Panel b) shows a typical situation in Test 3, in which two rays are traced from each particle and a $\Gamma^{\rm shld}$ is determined for each ray.  These are then combined to find the total shielded photoionisation rate for the particle (Eq.~\ref{eq:gammashld}).  In this case, radiation that should be incident on slab 2 from the direction of slab 3 is not accounted for by the solution $\mathcal{G}$ and vice versa.  Panel c) shows a 2-D cartoon of the situations in Test 4 in which rays are traced in all directions.  A component of the errors in these tests is due to an extension of the problem in using $\mathcal{G}$ described for panel b).  }
\label{fig:rayexplain}
\end{figure}

The first variable, $\Gamma^{\rm far} / \nH$, is the ratio of the amplitude of the incident radiation field over the density of the slab, the second, 
$\tau^{\rm far}$, determines the incident spectrum (i.e. how much the user supplied spectrum has been hardened before entering the slab), the third, $T$, determines recombination and collisional ionization rates, and the fourth, $\tau^{\rm near}_{\rm H}$ is the thickness of the slab.  In both the analytic and the lookup table case, we label this solution  $\mathcal{G}$ and note that the values of $\tau^{\rm far}$, $\Gamma^{\rm far}$,  and $\tau^{\rm near}_{\rm H}$ are different for each of the $N_{\rm ray}$ rays traced from particle $i$.  In summary, the total shielded photoionisation rate $\Gamma^{\rm shld}$ for particle $i$ is computed in the following way.  We trace rays $k=1,\cdots,N_{\rm ray}$ and calculate,
\begin{eqnarray}
  \label{eq:gammafar}
  \Gamma^{\rm far}_k   &=& \int_{\nuth}^{q\nuth} 
  \frac{\Inu \sigma}{h\nu}  e^{-\tau^{\rm far}_k} d\nu  
\\
  \label{eq:gammashldk}
  \Gamma^{\rm shld}_k &=& 
  \mathcal{G} \left(
    \frac{\Gamma^{\rm far}_k}{\nion{H}},
    T,
    \tau^{\rm far}_{k}, 
    \tau^{\rm near}_{{\rm H},k}
  \right)
\\
  \label{eq:gammashld}
  \Gamma^{\rm shld} &=&
  \sum_{k=1}^{\rm N_{\rm ray}} \frac{4 \pi}{\rm N_{{\rm ray}}} 
  \Gamma^{\rm shld}_k\,.
\end{eqnarray}
This whole process is illustrated in Fig.~\ref{fig:rayexplain}.  The determination of $\Gamma^{\rm shld}$ has no dependence on the neutral fraction of particle $i$ or any of its neighbors and therefore the solution is numerically stable. However, it requires extra computational effort to evaluate the function $\mathcal{G}$ and introduces errors due to the lack of interaction between different {\em near} slabs.  In the following section we will quantify these errors. 

\section{Tests and Verification}

In this section, we present several tests performed in order to validate \urchin.  We begin with simple test cases with a known analytic solution, and end with more realistic tests that involve gaseous galactic halos drawn from a cosmological simulation.  For those tests in which an analytic solution is not available, we compare the \urchin\ results to those of a straight-forward numerical solver which we will call ${\mathcal{NS}}$ to distinguish it from \urchin.  After we have verified ${\mathcal{NS}}$ against analytic solutions, we will simply refer to ${\mathcal{NS}}$ as the analytic solution. 

\subsection{Test 1: Analytic Slab Solution}
\label{sect:analytic}

\begin{figure}
\begin{center}
\includegraphics[width=0.45\textwidth]{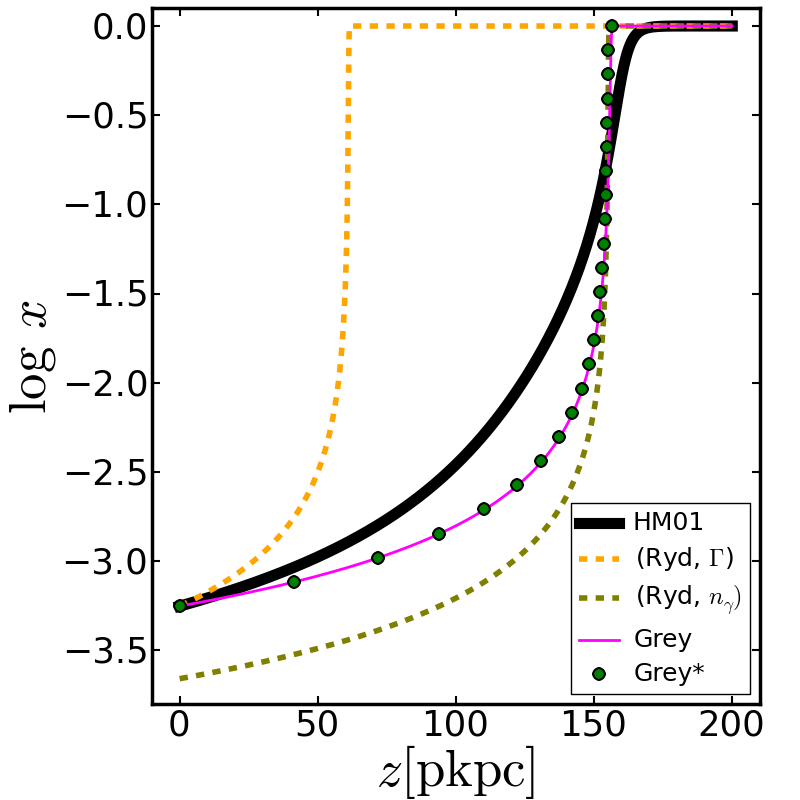}
\end{center}
\caption{Test 1: Equilibrium neutral fraction $x$ as a function of depth $z$ in the case of plane-parallel radiation incident onto a slab of hydrogen with constant and uniform density $\nHslab$ and temperature $T=10^4 {\rm K}$.  The {\em thick black line} is the solution when using the spectrum of Haardt \& Madau 2001 restricted to photons with energy $1 < h\nu / h\nuth < 4$ ($\Inu^{\rm HM}$).  The {\em pink line} is the ${\mathcal{NS}}$ result for the case of monochromatic photons with energy $h\nu=19.2$~eV (the grey approximation of $\Inu^{\rm HM}$); green dots are the corresponding analytic result discussed in the Appendix. The excellent agreement demonstrates the accuracy of ${\mathcal{NS}}$.  By construction, the grey approximation reproduces both the number density of photons, $n_{\gamma}$, and the photoionisation rate, $\Gamma$, of $\Inu^{\rm HM}$ in the optically thin limit. The {\em dashed yellow} and {\em dashed olive} curves represent cases in which monochromatic photons with $h\nu=$~13.6~eV were incident and the flux was normalised to reproduce either $\Gamma$ or $n_{\gamma}$, respectively.  These monochromatic approximations to HM01 give inaccurate results either in the optically thin limit, and/or in the position of the ionisation front, illustrating the need to properly sample the input spectrum.}
\label{fig:slab}
\end{figure}

This test involves plane-parallel radiation with flux $F$ incident from one side onto a slab of hydrogen gas of thickness $L_{\rm slab}$, uniform density $n_{\rm H}$, and fixed uniform temperature $T_{\rm slab}$.  The surface of the slab is coincident with the $z=0$ plane and the bulk extends in the $z>0$ direction.  In the case of monochromatic radiation, this problem has an analytic solution which we derive in Appendix A4.  The purpose of this test is to verify ${\mathcal{NS}}$ and to illustrate the dependence of the solution on the assumed spectrum of the incoming radiation.    

The equilibrium neutral fraction as a function of depth, $x(z)$, is obtained with ${\mathcal{NS}}$ by dividing the slab into many thin slices perpendicular to the $z$-axis.  Starting from $z=0$ and working downwards, we solve for $x$ in one slice at a time using the following algorithm: 1) determine the \ion{H}{I} optical depth, $\tau = \int_0^z x \, \nH dz$, above the current slice; 2) calculate an attenuated photoionisation rate in the slice, $\Gamma = \int_{\nuth}^{\infty} F \sigma \exp(-\tau) d\nu$; 3) determine $x$ in the slice by plugging $\Gamma$ into an analytic solution (Eq.~\ref{eq:xroots}).  To avoid errors due to finite slice width, we choose the number of slices such that each has a total hydrogen optical depth $\tauH = \NH \sigma_{\rm th}$ below unity.  This guarantees they will always be optically thin when considered individually. The numerical values for the parameters of this test are, $L_{\rm slab} = 200$~pkpc, $\nH = 1.5 \times 10^{-3} {\rm cm^{-3}}$ (500 times the cosmic mean $\nH$ at redshift 3), and $T_{\rm slab}=10^4$~K.  We compare the analytical solution to that obtained by ${\mathcal{NS}}$ in Fig.~\ref{fig:slab} (pink line labelled {\em grey} versus green symbols): they agree very well.

An advantage of the reverse ray tracing approach is the high fidelity with which the spectrum of the ionising background can be sampled.  This is non-trivial in forward ray tracing schemes and many rely on a simplified spectrum \cite[see discussion in][]{Mirocha_12}. We can use ${\mathcal{NS}}$ to quantify the accuracy of such approximations. The slab parameters were chosen in order to produce a fully neutral region on the far side of the slab and a total column density $\NHI = 10^{20.3} {\rm cm^{-2}}$ when the redshift $z=3$ \citealt{HM01} (HM01) UV background is incident.  In what follows, we will refer to the specific intensity of the HM01 UV background at redshift 3 between the frequencies $\nuth$ and $4 \nuth$ as $\Inu^{\rm HM}$.  In Figure~\ref{fig:slab} we compare HM01 with three different monochromatic approximations:  {\em i)} 1 Rydberg photons with a flux that produces the same (optically thin) density of ionising photons as $\Inu^{\rm HM}$, {\em ii)} 1 Rydberg photons with a flux that produces the same (optically thin) photoionisation rate as $\Inu^{\rm HM}$, and {\em iii)} the grey approximation of $\Inu^{\rm HM}$ which reproduces both the number density of photons and photoionisation rate of $\Inu^{\rm HM}$.  Photons in the grey approximation have an energy of 19.2~eV.


In all cases the monochromatic solutions transition to fully neutral more abruptly than the true HM01 solution.  This is due to the photons having a fixed photoionisation cross-section.  The spread in frequencies in the HM01 spectrum smooths the transition from highly ionised to fully neutral. In the worst case, this can cause an error of several orders of magnitude in the neutral fraction.  The monochromatic spectrum normalised to the same photo-ionisation rate (yellow line) recovers the correct neutral fraction in the optically thin region but underestimates the depth of the ionised region by almost 100~pkpc. The monochromatic spectrum with the same number density of ionising photons (olive line) underestimates the neutral fraction in the optically thin region by $\sim 0.5$~dex, but recovers the location of the ionisation front to better than 10~kpc. The grey approximation (pink line) recovers the ionised fraction in the optically thin region and the location of the ionisation front, but still has a maximum error of $\sim 0.75$~dex.

The dependence of $x(z)$ on the spectrum of radiation illustrates the need for accurate multi-frequency treatments.  An advantage of \urchin\ is that it can treat arbitrary spectral shapes accurately with approximately 100 frequency bins. In addition, the full attenuated ionising spectrum at a particle position is known when neutral fractions are calculated. For completeness we examine several other frequently used approximations to the HM01 spectrum in Appendix \ref{sect:powerlaw}, and various fitting formula used to approximate the hydrogen photoionisation cross section in Appendix \ref{sect:cross}.

\subsection{Test 2: Uniform Slab - Radiation Incident from One Side}

\begin{figure}
  \begin{center}
    \includegraphics[width=0.45\textwidth]{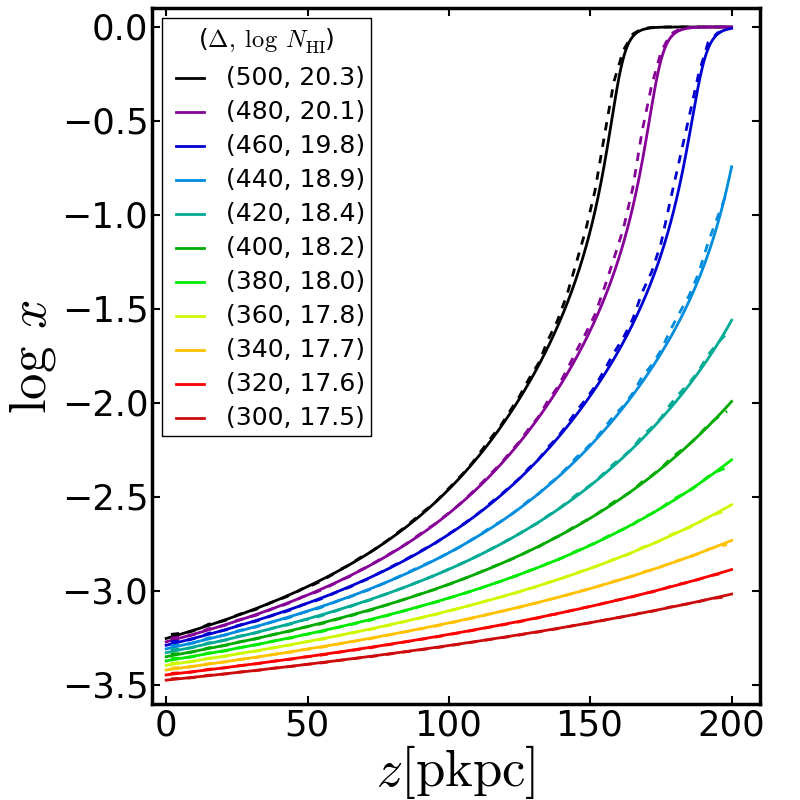}
    \includegraphics[width=0.45\textwidth]{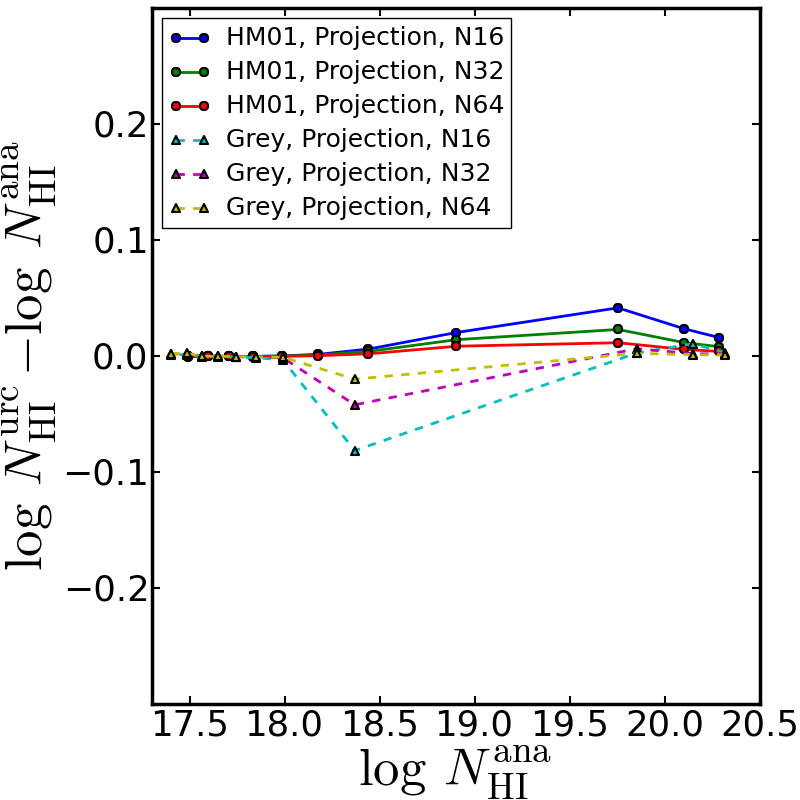}
  \end{center}
  \caption{ Test 2: Plane parallel radiation incident from one side onto 200~pkpc isothermal slabs of uniform density.  The spectrum is that of HM01 at redshift 3 integrated between 1 and 4 Rydbergs.  The slabs shown have $\nHslab \times$ \{1.0, 0.96, 0.92, 0.88, 0.84, 0.80, 0.76, 0.72, 0.68, 0.64, 0.60 \} (from black to red) which, at redshift 3, correspond to overdensities of $\Delta = $ \{500, 480, 460, 440,  420, 400, 380, 360, 340, 320, 300\}.  {\em Top panel}: equilibrium neutral fraction as a function of depth $x(z)$ for the analytic solution (solid lines) and the \urchin\ N16 resolution solution (dashed lines). Lines are labelled with the corresponding value of the over density ($\nH = \nH^{\rm mean} \times \Delta$), and the analytic neutral hydrogen column density through the slab, $\log N_{\rm HI}^{\rm ana}$.  {\em Bottom panel}: the difference in $\NHI$  between the \urchin\ and analytic solutions, as a function of the analytic $\NHI$.  Shown are the results when the HM01 spectrum is used (solid lines) and when the grey approximation is used (dashed lines) , with different colours referring to different particle resolutions.  In all cases, the \urchin\ reverse ray tracing solution is within 0.1~dex of the analytic value, with the error depending weakly on particle resolution.}
\label{test_two_plot}
\end{figure}

In this test we compare \urchin\ solutions to those of ${\mathcal{NS}}$.  To this end, we create a set of uniform slabs with the same geometry as in Test 1, but vary the volume densities $\nion{H}$ such that the projected HI column densities through the slabs cover the range $17.5 < \NHI / {\rm cm^{-2}} < 20.3$ (i.e. the range over which self-shielding becomes important).  To create SPH realizations of the uniform density fields we generate glass-like distributions with $16^3$, $32^3$, and $64^3$ particles (hereafter labeled N16, N32, and N64).  To model the plane parallel radiation in \urchin, we trace a single ray of length $L_{\rm slab}$ from each particle towards the surface of the slab. We calculate column densities by projecting all SPH particles onto a plane and measure the mean column on a fine grid of 2048$^2$ pixels.  Similar projections were used in \cite{Altay_2011}. 

In the top panel of Figure \ref{test_two_plot}, we compare ${\mathcal{NS}}$ solutions (solid lines) to those produced by \urchin\ (dashed lines) for slabs with different densities (different colours).  The \urchin\ solutions are from the lowest (N16) resolution SPH density fields. \urchin\ faithfully follows the dependence of neutral fraction on depth for all models, including those where the gas becomes mostly neutral. The bottom panel of the figure quantifies the errors in neutral hydrogen column densities calculated from the SPH realizations (solid lines).  Errors are below 0.05~dex at low resolution (N16, blue line), and improve with increasing resolution. The dashed lines quantify the \urchin\ errors in the case of monochromatic radiation, for which the analytic solution $\mathcal{G}$ does not require the construction of the interpolation table discussed in Section~\ref{sect:partupdate}. The errors are slightly larger here as the ionisation front becomes very steep when the radiation is monochromatic.  However, this test demonstrates that our method of splitting the optical depth into a contribution from near and far particles works well for single ray applications. 

\subsection{Test 3: Uniform Slab - Radiation Incident from Two Sides}

\begin{figure}
  \begin{center}
    \includegraphics[width=0.45\textwidth]{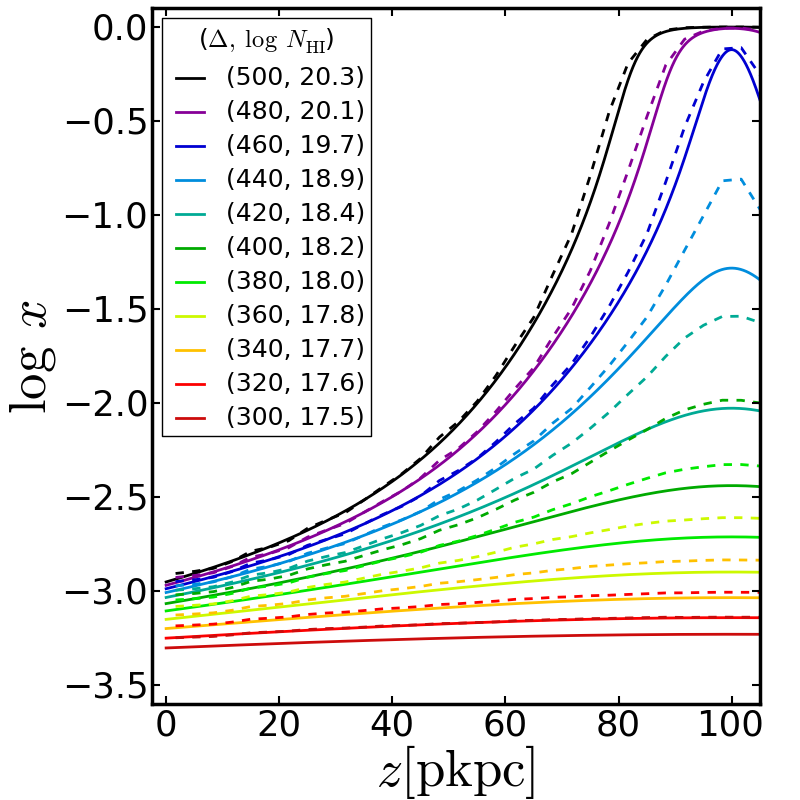}
    \includegraphics[width=0.45\textwidth]{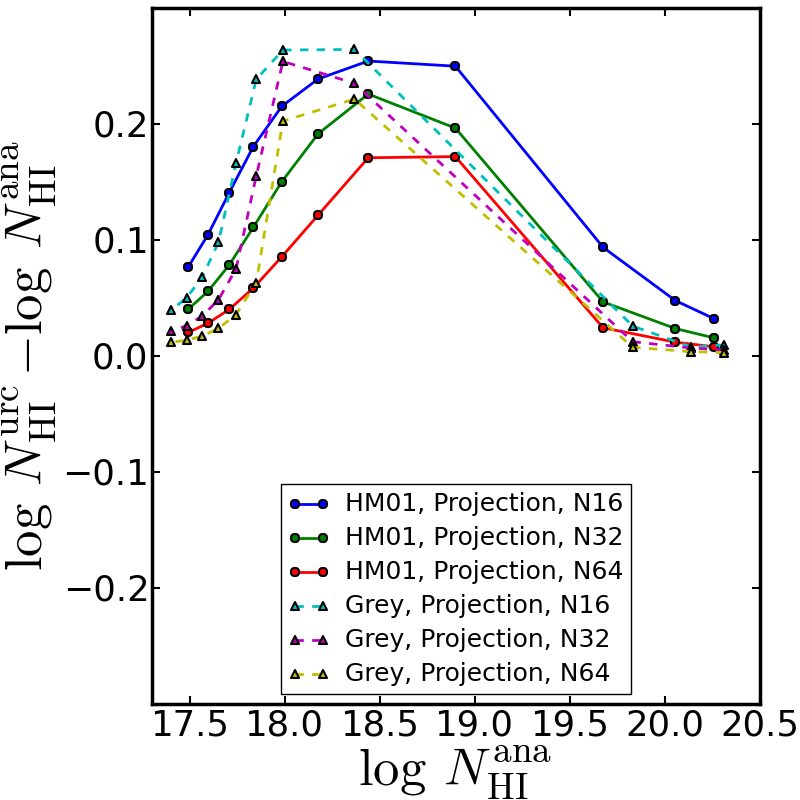}
  \end{center}
  \caption{ Test 3: Plane parallel radiation incident from two sides onto 200~pkpc isothermal slabs of uniform density. The spectrum is that of HM01 at redshift 3 integrated between 1 and 4 Rydbergs.  The panels are arranged as in Figure \ref{test_two_plot}. {\em Top panel}: for slabs with $\Delta \le 340$ or $\Delta \ge 460$, errors in the ionisation profiles are similar to Test 2.  In the intermediate regime both rays are important in determining the photoionisation rate at a particle and the assumptions made in the solution  $\mathcal{G}$ are not valid (see Figure~\ref{fig:rayexplain}). {\em Bottom panel}: the difference in $\NHI$  between the \urchin\ and analytic solutions, as a function of the analytic $\NHI$.  Because the importance of $\mathcal{G}$ is reduced at higher resolutions, these errors scale more strongly with particle resolution than those in Test 2.}
\label{test_three_plot}
\end{figure}

This test is identical to Test 2, except we irradiate the slabs from both sides.  To obtain the analytic solution, we modify $\mathcal{NS}$ as follows.  First we initialise the neutral fractions of all slab slices to their optically thin values, then we loop over the slices calculating the optical depth both above and below a given slice.  These optical depths are used to calculate a photoionisation rate and hence a new value for the neutral fraction.  We continue iterating over the slices until the neutral fraction in each slice has converged to one part in ten thousand. The solution is symmetric with respect to the centre of the slab at $z=100$~pkpc.  To model the plane parallel radiation in \urchin, we trace two rays from each particle, one in the $+z$ direction and one in the $-z$ direction. 

As in the previous test, we compare the \urchin\ (dashed lines) and analytic  (solid lines) solutions for the neutral fraction in the top panel of Fig.~\ref{test_three_plot}.  The goal of this test is to examine the accuracy of the {\em near} / {\em far} split described in section 4.1.3 when multiple rays are being used (diagramed in panel b of Figure \ref{fig:rayexplain}).  The algorithm introduces errors in the calculation of $\Gamma^{\rm shld}$ because each ray is considered independently.  To illustrate this point we will consider the process of calculating $\Gamma^{\rm shld} = 2\pi ( \Gamma_+^{\rm shld} + \Gamma_-^{\rm shld} )$ for a particle situated in the middle of a slab in which the equilibrium neutral fraction doen not form a neutral core (i.e. any slab with $\Delta < 460$).  First the $+z$ ray is traced and a $\Gamma^{\rm far}_+$ is calculated as in Eq. (\ref{eq:gammafar}).  This is then used as input to calculate $\Gamma^{\rm shld}_+$ as in Eq.(\ref{eq:gammashldk}) which represents the photoionisation rate at the bottom of the {\em near} slab.  The error occurs due to the fact that the {\em near} slab should also be irradiated from the $-z$ direction as all of the gas is highly ionised.  This leads to an overestimate of the opacity of each of the {\em near} slabs and in turn to an overestimate of the neutral fraction in the slab.  The errors are most severe during the transition from optically thin slabs to slabs that form a neutral core.  In slabs that do form a neutral core the error is absent as at least one ray always encounters a high optical depth. 

The resulting errors on the column density through the slab are shown in the bottom panel of Fig.~\ref{test_three_plot}. Different colours refer to different numerical resolutions, with solid lines representing the HM01 radiation field, and dashed lines the grey approximation. The \urchin\ optical depth is within 0.1 dex of the analytic result for columns $\NHI < 10^{18}$~cm$^{-2}$ or $\NHI > 10^{19.5}$~cm$^{-2}$ for the highest resolution slab. These column density limits correspond to the  cases where the slab is either mostly ionised everywhere, or develops a neutral core. In the intermediate regime, \urchin\ overestimates the neutral hydrogen column density by up to 0.15 dex at N64 resolution, and 0.25 dex at N16 resolution. The errors in the case of monochromatic radiation (dashed lines) are not significantly different, showing that the \urchin\ error is due to the {\em near}/{\em far} split, rather than the implementation of the look-up table for polychromatic spectra employed in ${\mathcal G}$ (as opposed to the analytic solution used for monochromatic spectra).

\subsection{Test 4: Galactic Halos - Uniform UV Background}

In previous sections we applied \urchin\ to very simple slab geometries for which we could calculate accurate analytic solutions. In this section we concentrate on the more realistic case of galactic halos.  In particular we focus on halos extracted from the \owls\ suite of cosmological galaxy formation simulations \citep{Schaye_2010}. This suite consists of a reference model ({\sc ref}) and more than 50 variations around that reference model which explore changes to sub-grid physics and other parameters.  The {\sc ref} model, which we make use of here, included a model for pressure in the numerically unresolved cold interstellar medium, star formation, the timed release of 11 chemical elements by type I and type II supernovae and AGB stars, radiative cooling due to the same elements in the presence of the HM01 ionising background, and energetic feedback from supernovae \citep{DallaVecchia_08, Schaye_08, Wiersma_09_cool, Wiersma_09_chem}. 

\subsubsection{Stacking Halos}

\begin{figure*}
  \begin{center}    
    \includegraphics[width=0.95\textwidth]{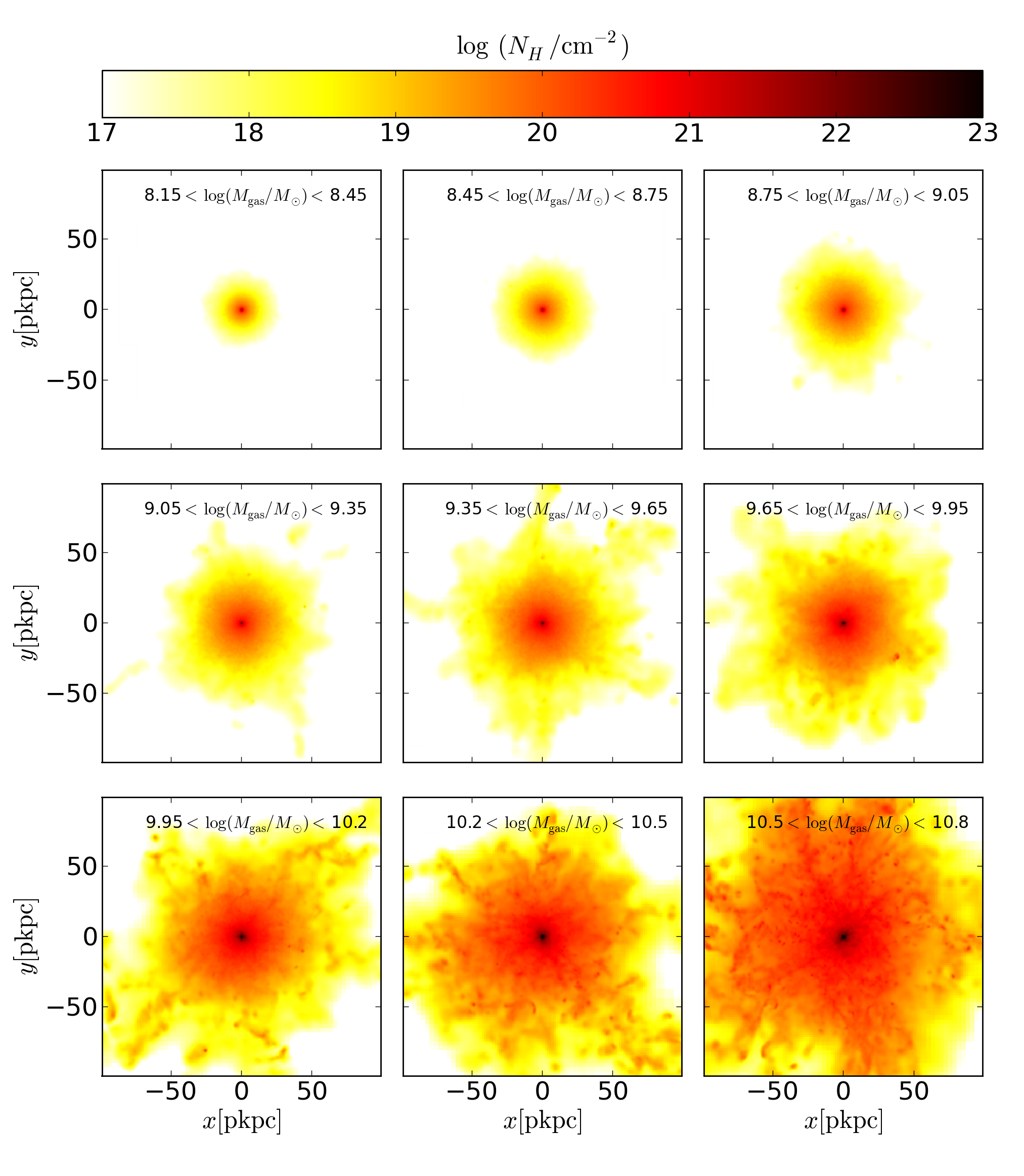}
  \end{center}
  \caption{ Test 4: Projected total hydrogen density for the nine halo mass bins used for constructing the halo stacks.  The color scale logarithmically covers the range between $\NH = 10^{17} {\rm cm^{-2}}$ and $\NH = 10^{23} {\rm cm^{-2}}$.  Each image shows a region 200 pkpc on a side and deep enough to include every particle in the stack.  These images give a sense of how the higher mass halo stacks are less spherically symmetric than the lower mass stacks, as there are fewer of them within the simulation volume to average over. }
\label{fig:stack_pics}
\end{figure*}

To identify these halos we first calculate a friends-of-friends group catalogue using the standard algorithm of \cite{Davis85} run on the dark matter particles, and linking baryonic particles to their nearest dark matter particle.  We then identify bound sub-structures (i.e. halos) within these FoF groups using the subfind algorithm \citep{Springel_01,Dolag_09} and associate these halos with galaxies.  Our goal is to create objects that have approximate spherical symmetry -- so we can calculate self-shielding analytically -- yet are representative of typical galaxy density profiles encountered in simulations.  To this end, we \lq stack\rq\ haloes of similar mass. First, haloes are grouped into mass bins according to their total gas mass, with the lowest mass bin edge being $M_{\rm gas} = 10^{8.15} \Msunh$ (100 gas particles) and the bin width equal to 0.3~dex. We then centre all haloes in a given bin on the location of the most bound particle of that halo. The mass of each particle in a stack is then adjusted such that the sum of all particle masses in the stack is at the logarithmic center of the mass bin.  Finally, the smoothing lengths and densities of all particles in a stack are recalculated.  These stacks, while still containing some particle noise, are now (nearly) spherically symmetric. 
In Figure~\ref{fig:stack_pics} we show images of each halo stack out to a radius of 100 pkpc.  The color scale logarithmically covers the range between $\NH = 10^{17} {\rm cm^{-2}}$ and $\NH = 10^{23} {\rm cm^{-2}}$.  We show this figure mainly to give the reader a visual impression of how the halo stacks become less spherically symmetric in the higher mass bins due to the smaller number of halos in each bin.

\begin{figure*}
  \begin{center}    
    \includegraphics[width=0.95\textwidth]{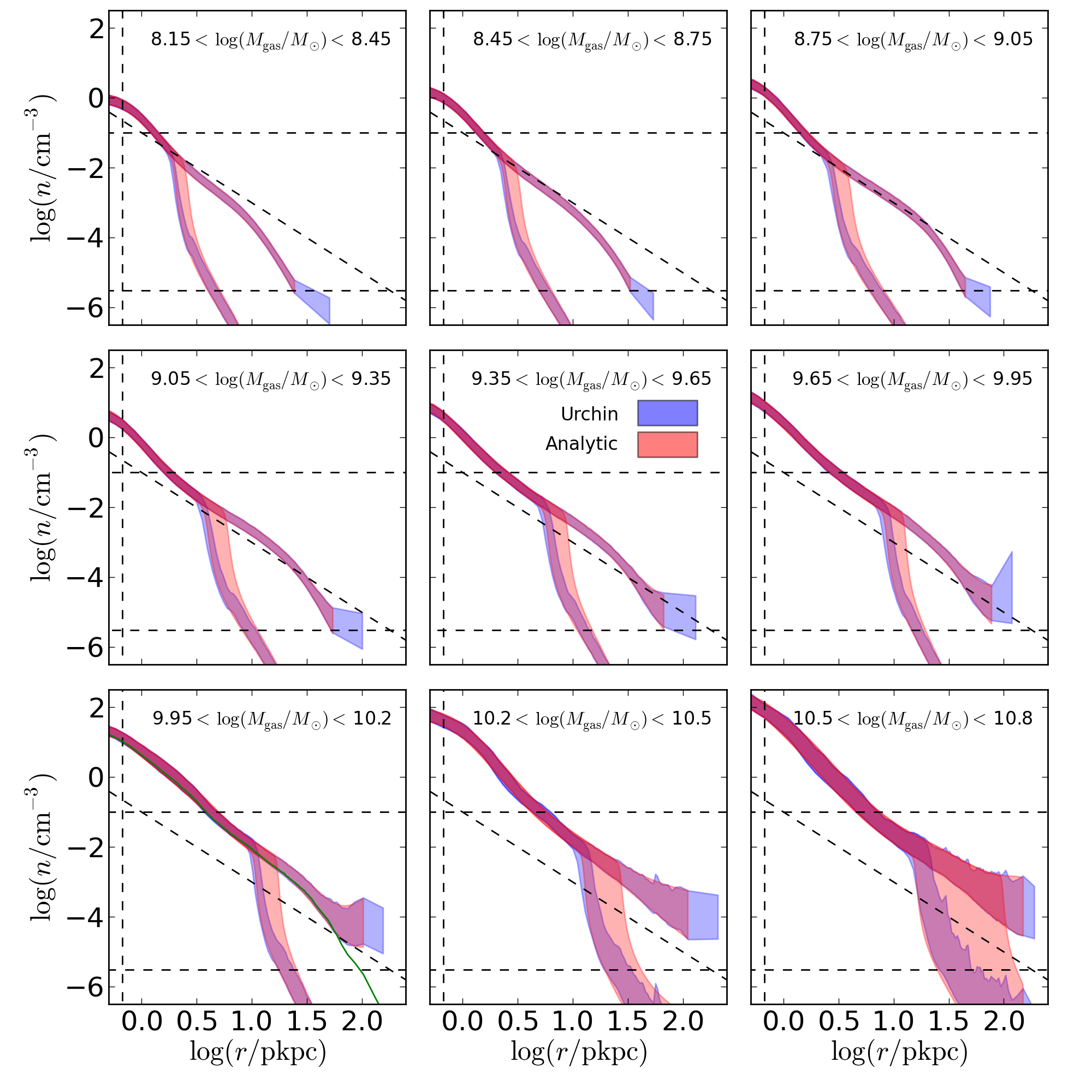}
  \end{center}
  \caption{ Test 4: $\nH$ and $\nHI$ radial profiles for \owls\ halo stacks irradiated with a uniform UV-background. Different panels correspond to different halo gas mass ranges as indicated in each panel.  The lower horizontal dashed line indicates the cosmic mean value of $\nH$ at $z=3$, the upper horizontal line indicates the star-formation density threshold used in the {\sc ref} \owls model, the vertical line indicates the gravitational softening length, and the diagonal line shows an arbitrarily normalised $r^{-2}$ relationship to guide the eye.  Both the total ($\nH$) and neutral ($\nHI$) hydrogen number density are shown with shaded regions. 
{ 
The two quantities are equal at small radii but diverge when $\nH \approx 10^{-2.0}$ cm$^{-3}$.  The total hydrogen profile continues to scale as roughly $r^{-2}$ but the neutral hydrogen profile becomes much steeper causing the shaded regions to have a shape similar to the greek letter lambda.  
}
The SPH density fields on which \urchin\ was run are shown in blue while our analytic solutions are shown in red.  All SPH profiles (blue) are constructed from mass weighted averages at a given radii.  The width of the shaded regions indicates one sigma variations from the median, and hence are a measure of deviations from spherical symmetry in the stacks.   In the lower left panel we also indicate the volume weighted average of $\nH$ with a green line.  The \urchin\ solution and the analytic solution always overlap.  The two solutions are most different in regions where the $\nHI$ profile is steepest. }
\label{fig:lambda}
\end{figure*}

\subsubsection{Radial Profiles}
In Fig.~(\ref{fig:lambda}) we plot radial density profiles of the stacked haloes.  In each panel there are four black dashed lines.  The  lower horizontal line indicates the cosmic mean value of $\nH$ at $z=3$, the upper horizontal line indicates the star-formation density threshold used in {\sc ref}, the vertical line indicates the gravitational softening length, and the diagonal line is an arbitrarily normalised $1/r^2$ line to guide the eye.  Both the total and neutral hydrogen number density are shown with shaded regions.  The two quantities are equal at small radii causing the shaded regions to have a shape similar to the greek letter lambda in each panel. 

\subsubsection{Total Hydrogen Radial Profiles}

Total hydrogen density profiles were calculated from the SPH stacks and are plotted as the upper blue shaded regions. The width of the regions correspond to one sigma variations from the mean density at a given radius.  This is a measure of the deviations from spherical symmetry shown in Figure~\ref{fig:stack_pics}.  The shaded regions are mass weighted averages but for reference we
also plot the volume weighted average, i.e. the sum of all particle masses in a radial shell divided by the volume of the shell, with a green line in the bottom left panel.  The volume weighted density hugs the lower end of the mass weighted density.  In an SPH distribution with no particle noise the two quantities would be equal; however, any clumping will increase the mass weighted quantity relative to the volume weighted quantity.  The offset between the two at small radii indicates the level of particle noise in the spherically symmetric part of the SPH stacks while the differences at large radii are due to clumps caused by stacking a finite number of halos.  Note that this noise was mostly absent in the previous tests due to the use of glass like distributions.  

For each mass bin we construct two smooth analytic $\nH$ profiles by fitting a polynomial through the one sigma variations described above (upper red shaded region).  The differences between the two are only visible at the larger radii of the more massive bins.  We use the point where the volume weighted $\nH$ intersects the cosmic mean $\nH$ to define a radius for each analytic profile which is why the red shaded regions are truncated at smaller radii than the blue.  This provides us with a perfectly smooth and spherically symmetric approximation to our SPH stacks.  In general, the density at a given radius is monotonically increasing with halo mass as is the radius of the halos.  Each halo has a profile close to $1/r^2$ at intermediate radii but becomes steeper at both smaller and larger radii.

\subsubsection{Neutral Hydrogen Radial Profiles}
To calculate neutral hydrogen profiles in spherically symmetric gas with $\mathcal{NS}$ we did the following.  Each halo is divided into $N_{\rm sh}$ shells.  From each shell, we trace $N_{\theta}$ rays which sample the azimuthal angle between $0$ and $\pi$ radians.  The photoionisation rate in each shell is calculated by summing the contribution from each ray and is then used to calculate a new neutral fraction. We loop over all shells and iterate until the neutral fraction in each shell has converged to one part in ten thousand.  We find that our results are numerically converged when using $N_{\rm sh}=1600$ and $N_{\theta}=13$.  We use $\mathcal{NS}$ to calculate $\nHI$ profiles from the $-1~\sigma$ and $+1\sigma$ $\nH$ profiles, and show the results as the lower red shaded regions. 

We also use \urchin\ to calculate the neutral fraction of every particle in each stack and construct average $\nHI$ profiles in the same way as we calculated the $\nH$ SPH density profiles. These are shown as the lower blue shaded areas.  When irradiated from outside, each halo develops a characteristic ionisation structure with a neutral core, a sharp ionization front, and an optically thin region.  In the core, $\nH = \nHI$.  When the density drops to $\log \nH \approx -2$ the neutral core transitions to a steep ionisation front.  At densities between $\log \nH = -4$ and $\log \nH = -5$ the ionisation profiles transition into the optically thin regime in which the neutral fraction is proportional to the density, $x = \alpha \nH / \Gamma$ . In all cases, the \urchin\ solution overlaps with the analytic solution.  The differences between the two are largest in the ionization front but match the analytic solution in the optically thin and optically thick regimes.   This is the spherical corrolary to the situation in Test 3 in which the errors were largest for intermediate column density slabs.  In a spherical geometry, even with a neutral core, some rays will sample the problematic columns between $10^{18} < \NHI/{\rm cm}^{-2} < 10^{19.5} {\rm cm}^{-2}$.

\subsubsection{Neutral Hydrogen Column Density Profiles}

\begin{figure*}
  \begin{center}    
    \includegraphics[width=0.95\textwidth]{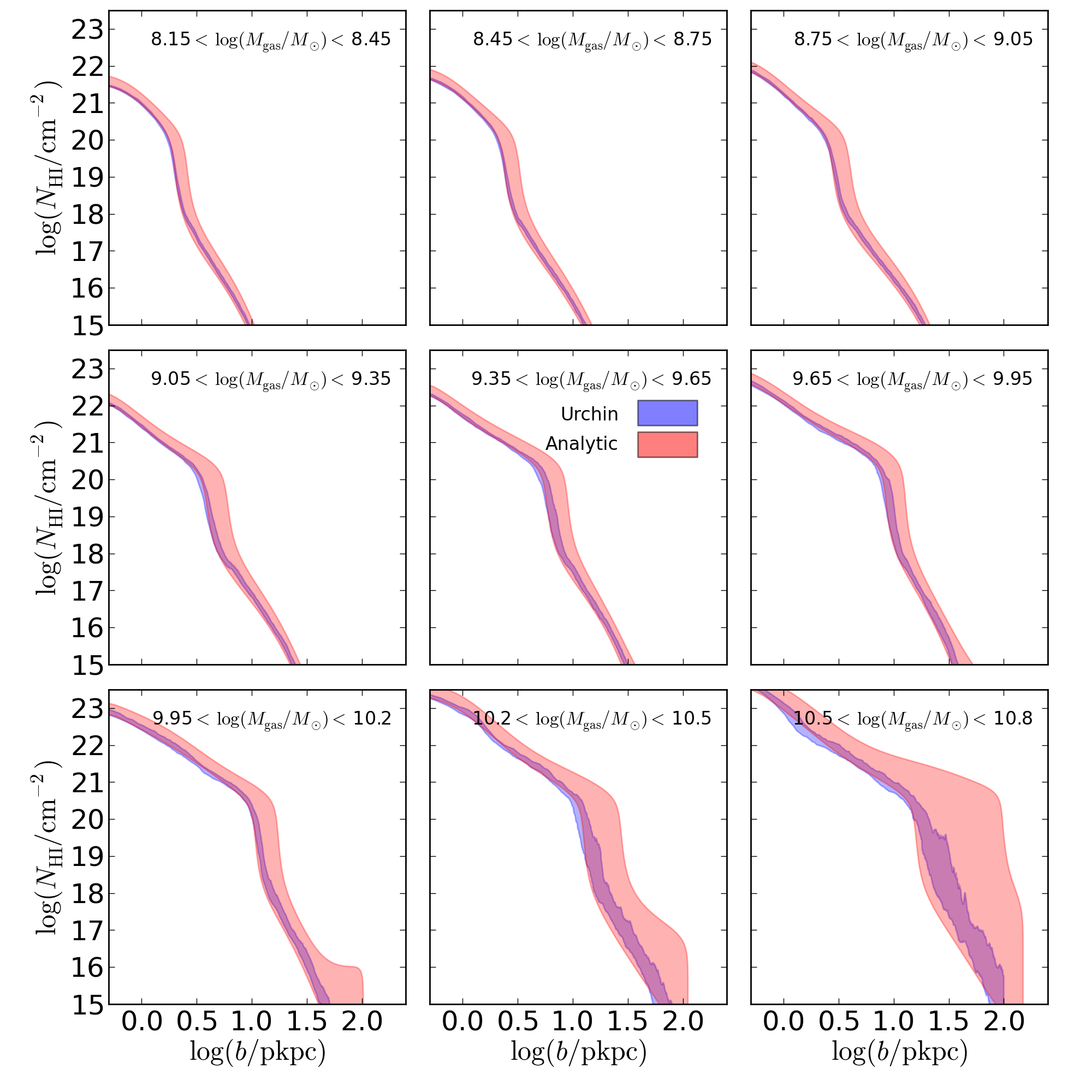}
  \end{center}
  \caption{ Test 4: $\NHI$ impact parameter profiles for \owls\ halo stacks irradiated with a uniform UV-background. Different panels correspond to different halo gas mass ranges as indicated in each panel.  The SPH density fields on which \urchin\ was run are shown in blue while our analytic solutions are shown in red.  The SPH profiles (blue) are constructed by projecting all particles in a halo stack onto a plane.  The width of the shaded regions indicates one sigma variations from the median value at a given impact parameter.  The \urchin\ solution and the analytic solution always overlap.  }
\label{fig:stack_NH1}
\end{figure*}

In Figure~\ref{fig:stack_NH1} we show the neutral hydrogen column density $\NHI$ as a function of impact parameter for the halo stacks. To calculate $\NHI$ profiles in the analytic case we simply integrate lines of sight through the spherically symmetric shells shown in Figure~\ref{fig:lambda} and tabulate the results as a function of impact parameter $b$.  We calculate these profiles for the $-1~\sigma$ and $+1\sigma$ spherical mass profiles, and shade the region between the two in red.  To calculate the same quantity from the SPH distributions we project all particles in a stack onto a plane and measure the column density on a fine grid of 2048$^2$ pixels.  We then bin these pixels in impact parameter and show the one sigma variation around the mean as a shaded blue region.  For density profiles of the form $\nH \propto r^{-n}$ with $n > 1/2$ the dominant contribution to $\NHI$ along a line of sight comes from the smallest radii.  Because of this, we can associate lines of sight in Figure~\ref{fig:stack_NH1} with the innermost radius they probe.  This gives rise to a corresponding region in the $\NHI(b)$ plots for each of the regions described in the previous section (neutral core, ionisation front, and optically thin outskirts).  In all mass bins, the \urchin\ solution overlaps with the analytic solution giving us confidence that \urchin\ can be used to accurately model neutral hydrogen absorbers in a cosmological context.

\section{Discussion and Conclusions}
\label{sect:discon}

We have presented and described a new publically available radiative transfer code called \urchin.  It is optimized to model the residual neutral hydrogen in the post-reionization Universe and relies on reverse ray tracing to avoid problems typically associated with modelling the UV background.  In particular, our implementation allows for the preservation of symmetry in the input radiation field, a completely uniform sampling of gas resolution elements, a high fidelity sampling of spectral features, and some optimisations not possible in forward ray tracing schemes.    

We have validated \urchin\ in four sets of tests that range from comparison to analytic solutions in simple slab geometries to solving for the ionization structure in gaseous galactic halos.  The errors discussed in \S5 have two root causes.  One is particle resolution, and the other is the multi-valued relationship between points in space and resolution elements in SPH.  The only solution to the first problem is to run simulations with more particles.  The second problem prompted us to introduce the {\em near} / {\em far} split and the analytic slab solution $\mathcal{G}$ discussed in \S4.  This solution solves the multi-valued problem but is restrictive in its use of plane parallel geometry and its treatment of each ray / {\em near} slab pair independently (i.e. without considering the other pairs).  It is possible that an approach that parameterises a spherical analytic solution using some average of the {\em near} slab optical depths and some average of the incoming radiation from all rays as parameters would be more sucessful.  In addition, the analytic solution presented in Eq.~\ref{analytic_slab_eqn} considers uniform density slabs although there is a density derivative in the fully general solution.  Estimating $d\nH/dr$ along each ray and using this as an extra parameter in $\mathcal{G}$ could also prove useful in improving the solution. 

In the current implementation, we have shown that solutions generated by \urchin\ reproduce analytic solutions in the case of density fields approximating gaseous galacitc halos.  
We view \urchin\ as a first step in a series of incremental improvements to the standard treatment of the UV background.  Future steps will include helium as in \cite{Altay_08}, proximity zones from internal point sources, an equilibrium heating/cooling model, and non-equilibrium corrections for ionisation and heating.  

\section*{Acknowledgments}
This research was supported in part by the National Science Foundation
under Grant No. NSF PHY11-25915. The calculations for this paper were performed on the ICC Cosmology Machine, which is part of the DiRAC Facility jointly funded by STFC, the Large Facilities Capital Fund of BIS, and Durham University.

\label{lastpage}

\bibliographystyle{mnras}	
\bibliography{biblio}	

\appendix

{

\section{\urchin\ Parameter Choices}

The main inputs expected from a user of \urchin\ are an optically thin spectrum and parameters determining the number of rays per particle, $N_{\rm ray}$ and the distance each ray is traced, $l_{\rm ray}$.  In some cases, the correct choices for these parameters are obvious, in others there are no unique correct choices.  We will examine a few examples to clarify the situation.  

Cases involving plane parallel radiation incident onto isolated gas slabs are in the first category.  The number of rays is fixed to one or two depending on if the radiation is incident from one or both sides.  In addition, the rays need to be at least as long as the slab in the direction they are traced. Once the rays have exited the slab, they will not add any additional optical depth and will therefore not change the results.  These choices are independent of the input spectrum.  For cases in which an isotropic optically thin background is incident on isolated halos, the ray length should again be equal to the linear extent of the object in question but the number of rays to trace for each particle is not obvious.  \urchin\ allows a user to select from the available Healpix resolutions with $N_{\rm pix} = 12 N_{\rm side}^2$ with $N_{\rm side}$ equal to a positive integer.  A larger number of resolution elements in the density field will require higher angular resolution to fairly sample the region surrounding each particle.  In general, the value of $N_{\rm ray}$ is entirely dependent on the resolution of the density field.

To date, \urchin\ has mostly been applied to the problem of modelling the post-reionization UV background in periodic cosmological simulations.  In this case, it is important to keep in mind that the input optically thin spectrum \cite[for example][]{HM_12} already includes attenuation on large scales.  What we wish to model is the local attenuation in overdense regions.  It is also instructive to consider what happens as $l_{\rm ray}$ is increased from zero to infinity.  For $l_{\rm ray} = 0$ there is no attenuation and the results are identical to the optically thin limit.  As $l_{\rm ray}$ is increased, the local environment of the particle is probed and the shielded photoionization rate begins to differ from that of the optically thin case in self-shielded regions. As $l_{\rm ray}$ is increased further, the ray begins to probe regions of the cosmological volume that are completely uncorrelated with the starting location.  At this point, the results are equivalent to lowering the normalization of the input optically thin spectrum.  

One possible strategy for producing converged results is the following.  Begin with rays which are short compared to the size of density field resolution elements.  Increase the ray length in steps which are some fraction of the mean free path for the grey approximation of the input spectrum.  In most realistic applications, the results will converge before the rays reach a sizeable fraction of the mean free path.  Next, test for convergence in $N_{\rm ray}$.  We note that these results will depend on the maximum frequency one includes in the input spectrum and on the particular statistic one is interested in.  Our suggested values of $l_{\rm ray} = 100$ pkpc and $N_{\rm ray}$=12 result from our application of the above algorithm to the \ion{H}{I} column density distribution function when truncating the UV background  spectrum at four Rydbergs \cite[see][]{Altay_2011}.  As with all numerical results, the surest way to have confidence in any answer produced by \urchin\ is to perform convergence tests.

}

\begin{figure}
  \begin{center}    
    \includegraphics[width=0.5\textwidth]{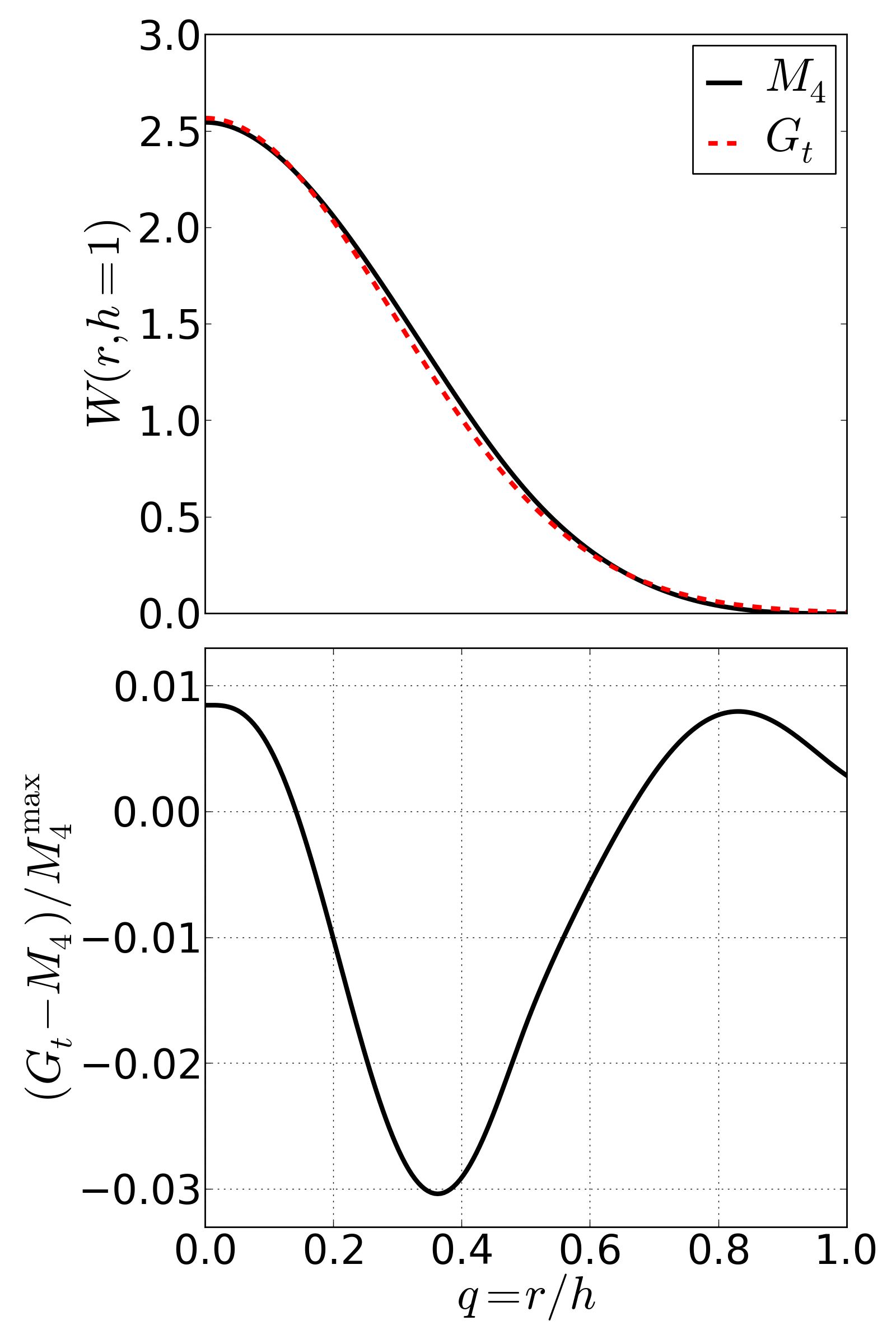}
    \end{center}
  \caption{
{\em Top panel}: The traditional $M_4$ spline (see Eq. \ref{eq:spline}) and the truncated gaussian used in \urchin\ (see Eq. \ref{eq:Gt}) for $h=1$. {\em Bottom panel}: The difference between the two smoothing kernels normalised by the maximum value of the $M_4$ kernel, $M_4^{\rm max}$.  This panel is independent of $h$. The normalisation of the kernels is identical, $4 \pi \int_0^h r^2 M_4 dr = 4 \pi \int_0^h r^2 G_t dr = 1$, and differences at specific radii are always smaller than 3\% of the maximum $M_4$ value.   }
\label{fig:Gt_kernel}
\end{figure}

{

\section{Comparison Between Smoothing Kernels}

In Fig. \ref{fig:Gt_kernel}, we compare the traditional $M_4$ spline (see Eq. \ref{eq:spline}) and the truncated gaussian used in \urchin\ (see Eq. \ref{eq:Gt}).  We note that the gaussian has dropped to less than 1\% of the maximum $M_4$ value when it is truncated.  The normalisation of the two kernels is identical by construction, $4 \pi \int_0^h r^2 M_4 dr = 4 \pi \int_0^h r^2 G_t dr = 1$ and their shapes are very similar with differences between the two always smaller than 3\% of the maximum $M_4$ value.    Using the truncated gaussian as opposed to the cubic spline has the effect of slightly increasing the contribution from the core and the wings of the kernel while slightly decreasing the contribution from intermediate radii.  While good SPH kernels posses certain desirable qualities, for example compact support, symmetry around $r=0$, and smoothness \citep[see][]{Price_05}, their specific shape is somewhat arbitrary.  

}

\section{Analytic Hydrogen Ionization Solutions}

\subsection{Photoionization Cross Section}
\label{sect:cross}

The photoionization cross section for hydrogenic atoms in the ground state can be expressed analytically as,
\begin{eqnarray}
\label{sigma_ana_eqn}
\sigma &=& \frac{\sigth}{Z^2} \left( \frac{\nuth}{\nu} \right)^4
\frac{\exp \left\{ 4 - [(4 \tan^{-1} \epsilon) / \epsilon] \right\} }
{1 - \exp(-2\pi/\epsilon)} \nonumber\\
\epsilon &=& \sqrt{ \frac{\nu}{\nuth}-1 }, \quad
\sigth = \frac{2^9 \pi}{3 e^4} \alpha_{\rm fs} \pi a_0^2\,, 
\end{eqnarray}
where $\alpha_{\rm fs}$ is the fine structure constant, $a_0$ is the Bohr radius, and $Z$ is the atomic number of the atom.  Two fitting formula are commonly used in the literature. The first is a power law form, usually accompanied by a citation to \cite{1989agna.book.....O},
\begin{eqnarray}
\label{sigma_pwr_eqn}
\sigma &=& \sigth \left( \frac{\nu}{\nuth} \right)^{-3}\,.
\end{eqnarray}
This simple form can be useful for analytic treatments, but is not as accurate as the fit due to \cite{1996ApJ...465..487V},
\begin{eqnarray}
\label{sigma_ver_eqn}
\sigma &=& \sigma_0 (x-1)^2 \, x^{0.5P-5.5} \left( 1 + \sqrt{x / x_a} \right) \nonumber\\
x &\equiv &\frac{h\nu}{E_0}\nonumber\\
E_0 &=& 0.4298\, {\rm eV}, \quad 
\sigma_0 = 5.475 \times 10^4 \, {\rm Mb} \nonumber \\
x_a &=& 32.88, \quad P = 2.963\,.
\end{eqnarray}
We explore the errors these approximations introduce into column density calculations in the next section.

\subsection{Power-law approximations to the Haardt \& Madau 2001 spectrum}
\label{sect:powerlaw}

\begin{figure*}
  \begin{center}
    \includegraphics[width=0.3\textwidth]{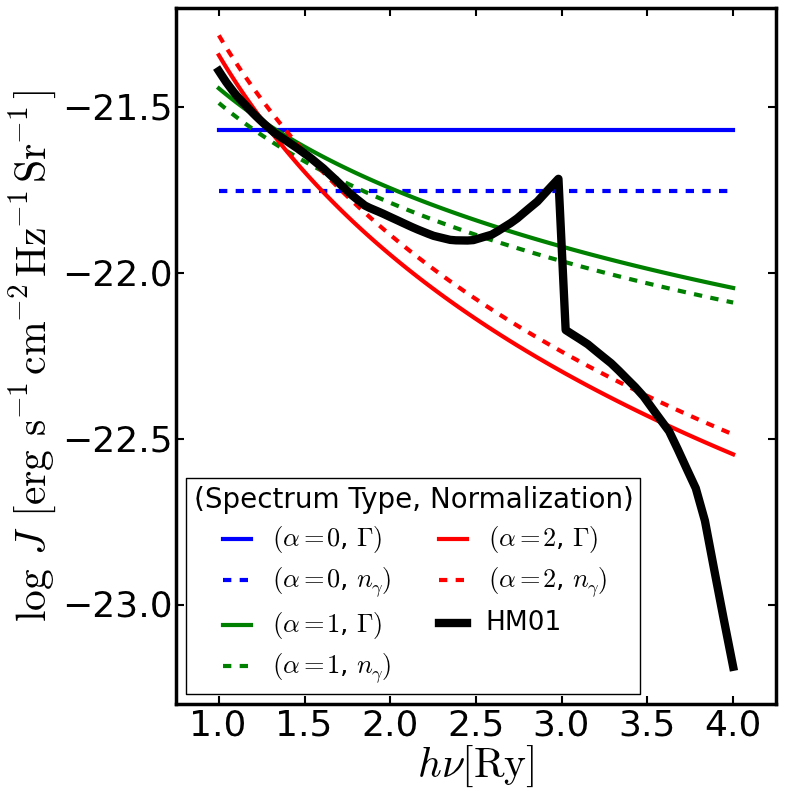}
    \includegraphics[width=0.3\textwidth]{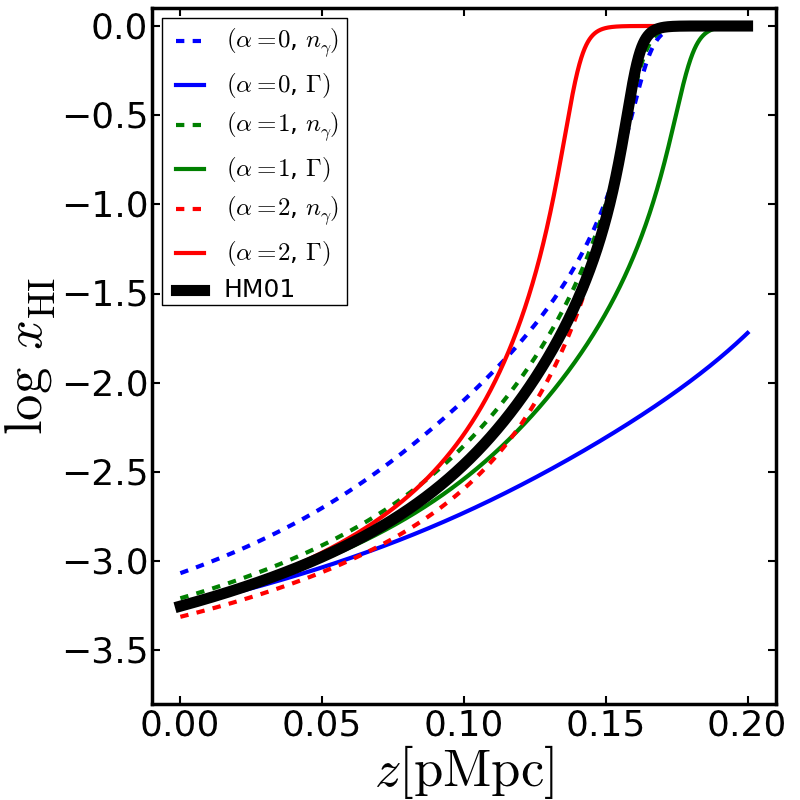}
    \includegraphics[width=0.3\textwidth]{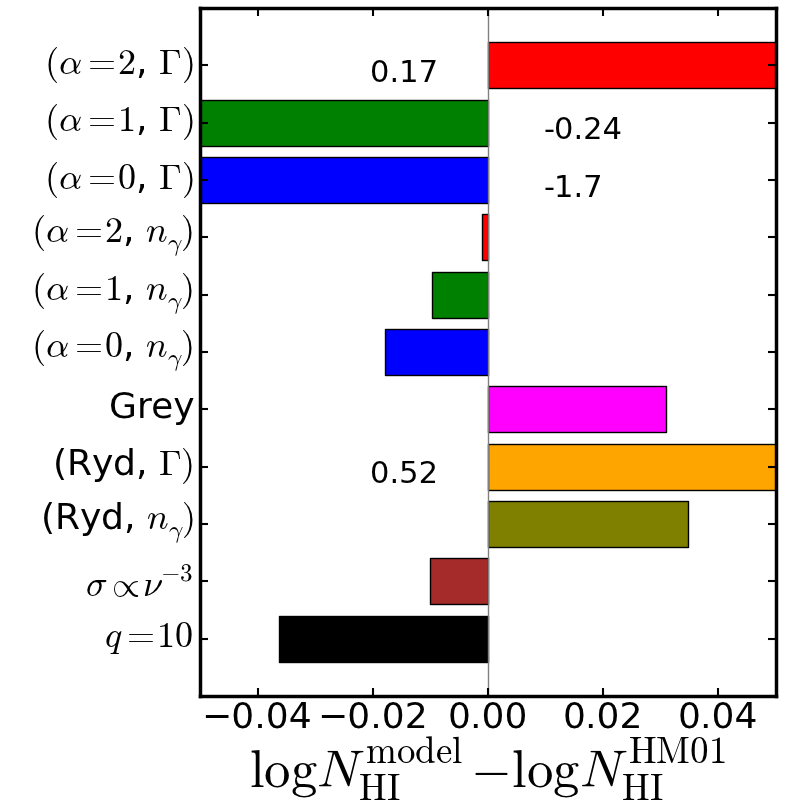}
    \end{center}
  \caption{
{\em Left panel}: The $z=3$ \protect\cite{HM01} UV-background (HM01) spectrum compared to three different power-law spectra,  $J(\nu)=J_0\,(\nu/\nu_{\rm th})^{-\alpha}$, for $\alpha=0,1,2$ shown blue, green and red, respectively.  For the solid lines, $J_0$ is chosen such that the power-law spectrum has the same photoionisation rate as the HM01 spectrum, for the dashed lines $J_0$ is chosen such that they have the same number density of ionising photons. The spike around 3 Rydbergs is due to Helium Lyman-$\alpha$ emission. 
{\em Middle panel}: equilibrium neutral fraction, $x_{\rm eq}(z)$ as a function of depth $z$ into a slab with uniform hydrogen density $\nHusphere$ (500 times the cosmic mean $n_{\rm H}$ at $z=3$) and $T=10^4$~K. The thick black line is for the full HM01 spectrum, blue, green and red are the corresponding power-law approximations. Such power-law models work relatively well, provided the amplitude $J_0$ of the spectrum is chosen such that the spectra have the same number density of ionising photons (dashed lines). 
{\em Right panel}:  difference between the total $\NHI$ for these spectral approximations and that in the HM01 model. For the four bars that go off the plot, we have indicated the horizontal value in the panel.  The error in $\NHI$ remains below 0.05~dex for all cases except those in which the  approximating spectra are normalized to have the same optically thin photoionization rate  in which case the errors can be much larger.  Of course the neutral fraction in the optically thin case is incorrect unless spectra are normalised to have the same optically thin photoionization rate. For reference, we also show as a brown bar differences in the hydrogen columns for the HM01 spectrum due to using a simple power law for the photoionisation cross section, as opposed to using the fit from \protect\cite{1996ApJ...465..487V}, see Eq.(\ref{sigma_ver_eqn}). Finally the black bar shows the effect of truncating the HM01 spectrum at 10~Rydberg, as opposed to 4~Rydberg.}
\label{test_one_plt}
\end{figure*}

The left panel in Figure~\ref{test_one_plt} compares the shape of the \cite{HM01} (HM01) spectrum to power-law approximations of the form $J(\nu)=J_0\,(\nu/\nu_{\rm th})^{-\alpha}$, for $\alpha=0,1,2$. The HM01 spectrum has features due to re-emission, such as the Helium Lyman-$\alpha$ emission at 3~Rydberg. The effect of using such approximate spectra as opposed
 to the full HM01 spectrum is illustrated in the middle panel. We calculated the equilibrium neutral fraction as function of depth, $x_{\rm eq}(z)$, for plane-parallel radiation entering a slab with uniform hydrogen density $\nHusphere$ (500 times the cosmic mean $n_{\rm H}$ at $z=3$) and $T=10^4$~K. Power-law approximations to HM01 work reasonably well, as long
 as the spectra are normalised to give the same number density of ionising photons as the orginal HM01 spectrum (dashed lines). Normalising the power-law spectra to the same photoisation rate does not work well (solid lines). The right panel in Figure~\ref{test_one_plt} compares the differences in neutral columns at $z=0.2$~pMpc, between the full HM01 spectrum and these powerlaw approximations. The largest error occurs when the slab fails to become fully neutral, as in the flat spectrum case ($\alpha=0$).  The monochromatic grey approximation used in Section \ref{sect:analytic} (pink bar) overestimates the central HI column by $\le 0.04$~dex. The bottom two bars show the effect of approximating the photoionisation cross-section by a simple power law versus using the more accurate fit from \cite{1996ApJ...465..487V}, and using the HM01 spectrum up to 10~Rydberg, as opposed to 4~Rydberg as we have done up to now.

\subsection{Time dependent solution of the neutral fraction}
\label{sect:xtsoltn}

The rate of change of the hydrogen neutral fraction $x \equiv \nHI/\nH$ is determined by the rate of photoionisation $\Gamma$, collisional ionisation $\gamma$, and recombination $\alpha$ as well as the number density of free electrons $\nel$. 
\begin{equation}
  \label{eq:dxdtpre}
  \frac{dx}{dt} = -(\Gamma + \gamma \, \nel) x + 
  \alpha \, \nel (1 - x)
\end{equation}
If we decompose the free electron number density into a contribution from ionised hydrogen and a contribution $y \nH$ from all heavier elements, we can write: 
\begin{equation}
  \nel= (1-x + y) \nH
\end{equation}
Substituting into the previous equation yields:
\begin{eqnarray}
  \label{eq:dxdt}
  \frac{dx}{dt} &=& -[\Gamma + \gamma (1-x+y)\nH] x + \nonumber \\
  &&   \alpha (1-x+y) \nH (1 - x)
\end{eqnarray}
Grouping terms in powers of $x$, we can write $dx/dt$ in the form of a Riccati equation:
\begin{eqnarray}
  \frac{dx}{dt} &=& Rx^2+Qx+P 
  \\
  R &\equiv& (\gamma + \alpha) \nH  \nonumber \\
  Q &\equiv& -\left[ \Gamma + (\gamma + 2\alpha) \nH + 
    (\gamma + \alpha) \nH y \right]  \nonumber \\
  &=& -\left[ \Gamma + \alpha \nH + 
    R (1+y) \right]  \nonumber \\
  P &\equiv& \alpha \nH (1+y) \,.
\end{eqnarray}
The roots of the quadratic term are 
\begin{eqnarray}
  \label{eq:xroots}
  x_{-} &=& {-Q - (Q^2-4PR)^{1/2} \over 2R}\\
  x_{+} &=& {-Q + (Q^2-4PR)^{1/2} \over 2R}\,.
\end{eqnarray}
To determine which of these roots is the physical equilibrium solution we consider the case of pure hydrogen ($y=0$) in the absence of radiation ($\Gamma=0$).  In this case,  
\begin{eqnarray}
  \label{eq:xphys}
  P  &=& \alpha \nH \\
  R  &=& (\alpha + \gamma) \nH \\
  -Q &=& R+P \\
  Q^2 - 4PR &=& (R-P)^2 \\
  x_{-} &=& {(R+P) - (R-P) \over 2R} = 
  \frac{\alpha}{\alpha+\gamma} = x_{\rm eq}
  \\
  x_{+} &=& {(R+P) + (R-P) \over 2R} = 1
\end{eqnarray}
The collisional ionisation and recombination rates depend on temperature $T$
\citep[see for example the fits in][]{Theuns_1998} which, in general, will change as $x$ changes. However in the case of constant $T$, $\nH$, $y$, and $\Gamma$, the coefficients $P$, $Q$ and $R$ are constants as well.  We can rewrite the derivative using Vieta's formula as,
\begin{equation}
  \frac{dx}{dt} = R \, (x-x_{\rm eq})(x-x_{+}) 
\end{equation}
and solve for the time-dependent solution by separation of variables 
\begin{equation}
  \frac{dx}{R \, (x-x_{\rm eq})(x-x_{+})} = dt
\end{equation}

\begin{eqnarray}
  x(t) &=& x_{\rm eq} + (x_0-x_{\rm eq}) 
  \frac{ (x_{+} - x_{\rm eq}) \, F }
  { (x_{+}-x_0) + (x_0 - x_{\rm eq}) F } \nonumber\\
  F(t) &\equiv& 
  \exp \left[ \frac{-(x_{+}-x_{\rm eq})t}{t_{\rm rc}} \right] \nonumber\\ 
  t_{\rm rc} &\equiv & {1 \over (\alpha+\gamma)\nH}
\end{eqnarray}
In a gas composed only of hydrogen and helium, $y$ is bound between $0$ and $(1-X)/(2X)$ where $X$ is the hydrogen mass fraction of the gas.  The solution $x(t)$ is fully determined once values for $T$, $\nH$, $y$, and $\Gamma$ are specified.  In addition, all of the dependence on the ionisation state of elements other than hydrogen is contained in the variable $y$.  For all of the tests performed in this paper we set $y=0$.

\subsection{Neutral fraction in a plane parallel slab}
\label{sect:slabsoltn}

In the case of plane parallel monochromatic radiation with a photon flux $F$ incident on a semi-infinite slab of hydrogen gas with constant density and temperature, the ionisation structure $x(z)$ can be calculated analytically.  We orient our coordinate system such that the surface of the slab is coincident with the $x-y$ plane and the positive $z$-axis extends into the slab.  The photoionisation rate at $z$ is $\Gamma(z) = F \sigma e^{-\tau(z)}  = \Gamma_0 e^{-\tau(z)}$, where $\Gamma_0$ is the photoionisation rate at $z=0$ and $\sigma$ is the photoionisation cross-section for the monochromatic radiation.
In equilibrium, Eq.~(\ref{eq:dxdt}) with $y=0$ implies: 
\begin{align}
\label{eq:slabone}
  \frac{\Gamma}{\Gamma_0} =  
  {\rm Exp} \left[ -\int_0^z \nH x \sigma ds \right] = 
  \frac{\nH \mathcal{X}}{\Gamma_0 x}
\end{align}
where we have defined $\mathcal{X} \equiv \alpha (1-x)^2 - \gamma(1-x) x$. Taking the logarithm of the last two terms and
differentiating with respect to $z$ gives:
\begin{align}
  \nH \sigma = - \frac{1}{x} \left\{
    \frac{1}{\nH} \frac{d\nH}{dz} + 
    \frac{d \ln \mathcal{X}}{dz} - 
    \frac{d \ln \Gamma_0 x}{dz}  \right\}
\end{align}
Setting the derivative of $\nH$ to zero and integrating both sides over the interval $[0,z]$ yields the depth $\NH \sigma = z \nH \sigma$ at which the neutral fraction is $x$.  We can write this inverse solution in terms of the equilibrium neutral fraction in the absence of radiation (i.e. in collisional ionisation equilibrium), $x_{\rm ce} \equiv \alpha / (\alpha + \gamma) = x(z\rightarrow\infty)$, and the value of $x$ at the surface of the slab $x(z=0) = x_0$.  The solution $z(x)$ is then:
\begin{eqnarray}
\label{analytic_slab_eqn}
z \nH \sigma = \left( \frac{1}{x_0} - \frac{1}{x}\right) &+&
\ln \left[ \frac{ x( 1-x_0) }{ x_0( 1-x ) } \right] + \nonumber \\
&& \frac{1}{x_{\rm ce}} 
\ln \left[ \frac{ x( x_{\rm ce}-x_0 ) }{ x_0( x_{\rm ce}-x ) } \right]
\end{eqnarray}
This allows $x(z)$ to be mapped out.  Once $x$ is known, a photoionisation rate can be determined using Eq.~\ref{eq:slabone}.  We use this solution to verify our numerical solver ($\mathcal{NS}$ in the text) and then use $\mathcal{NS}$ in more general (non-monochromatic)  cases to verify \urchin.  In addition, Eq. (\ref{analytic_slab_eqn}) forms the basis of the solution $\mathcal{G}$.


\end{document}